\documentclass[iop]{emulateapj}

\usepackage{comment}
\usepackage{enumerate}
\usepackage[T1]{fontenc}
\usepackage{pslatex}
\usepackage{epstopdf}
\usepackage{amsmath}
\usepackage{graphicx}

\slugcomment{Accepted to the Astrophysical Journal: January 10, 2014 issue}

\def\kms{\rm{km \ s^{-1}}}

\def\deg{^\circ }
\def\solar{_\odot }

\shorttitle{Transformation of Virgo Dwarf IC3418}
\shortauthors{Kenney et al.}

\begin{document}

\title{Transformation of a Virgo Cluster Dwarf Irregular Galaxy by Ram Pressure Stripping: IC3418 and its Fireballs}
\author {Jeffrey D. P. Kenney\altaffilmark{1},
Marla Geha\altaffilmark{1},
Pavel Jachym\altaffilmark{1,2},
Hugh H. Crowl\altaffilmark{3},
William Dague\altaffilmark{1},
Aeree Chung\altaffilmark{4},
Jacqueline van Gorkom\altaffilmark{5},
Bernd Vollmer\altaffilmark{6}
}
%
\altaffiltext{1}{\scriptsize
Yale University Astronomy Department,
P.O. Box 208101, New Haven, CT 06520-8101 USA jeff.kenney@yale.edu}

\altaffiltext{2}{\scriptsize
Astronomical Institute, Academy of Science of the Czech Republic, Bocni II 1401, 141 00 Prague, Czech Republic}

\altaffiltext{3}{\scriptsize
Bennington College, Bennington VT, USA}

\altaffiltext{4}{\scriptsize
Department of Astronomy and Yonsei University Observatory, Yonsei University, 120-749 Seoul, Republic of Korea}

\altaffiltext{5}{\scriptsize
Department of Astronomy, Columbia University, 550 West 120th Street, New York, NY 10027, USA}

\altaffiltext{6}{\scriptsize
CDS, Observatoire astronomique de Strasbourg, UMR7550, 11 rue de lÕuniversitŽ, 67000 Strasbourg, France}



%
\begin{abstract}

We present optical imaging and spectroscopy and HI imaging of 
the Virgo Cluster galaxy IC~3418, which is likely a "smoking 
gun" example of the transformation of a dwarf irregular into 
a dwarf elliptical galaxy by ram pressure stripping. IC~3418 has 
a spectacular 17 kpc length UV-bright tail comprised  of knots, 
head-tail, and linear stellar features. The only H$\alpha$ emission 
arises from a few HII regions in the tail, the 
brightest of which are at the heads of head-tail UV 
sources whose tails point toward the galaxy ("fireballs"). Several of 
the elongated tail sources have H$\alpha$ peaks outwardly offset by 
$\sim$80-150 pc from the UV peaks, suggesting that gas clumps
continue to accelerate through ram pressure, leaving behind streams of 
newly formed stars which have decoupled from the gas. Absorption 
line strengths, measured from Keck DEIMOS spectra, together with UV 
colors, show star formation stopped  300$\pm$100 Myr ago in the 
main body, and a strong starburst  occurred prior to quenching.
While neither H$\alpha$ nor HI emission are detected in the 
main body of the galaxy, we have detected 4$\times$10$^7$ M$\solar$ 
of HI from the tail with the VLA. The gas 
consumption timescale in the tail is relatively long, implying that 
most of the stripped gas does not form stars but 
joins the ICM. The velocities of tail HII regions, 
measured from Keck LRIS spectra, extend only a small fraction of 
the way to the cluster velocity, suggesting that star formation 
does not happen in more distant parts of the tail.
Stars in the outer tail have velocities exceeding the escape 
speed, but some in the inner tail should fall back into 
the galaxy, forming halo streams. One likely fallback stream 
is identified.
\end{abstract}


\section {Introduction}


Dwarf elliptical (dE)\footnote{
In this paper we will follow the majority of recent authors and refer to the more massive early type dwarf galaxies as dEs, and lower mass ones as dSphs. However some authors \citep{kb12} prefer the name Spheroidal (Sph) for more massive early type dwarfs, since they are
distinct systems from giant Ellipticals, and dEs (or Sphs) and dSphs seem to be part of the same family. dE can be interpreted as "early type dwarf".
}
galaxies are the dominant galaxy type (by number) in
clusters \citep{bts87}, although their
origin has not been well understood \citep{fb94,con01,geha03}.
Since gas-poor dEs are preferentially found in cluster centers, 
whereas gas-rich dwarf irregulars (dI)
are usually found in cluster outskirts and outside of clusters \citep{bts87}, 
a natural origin for dEs is via gas stripping of dIs
as they fall into dense ICM (intracluster medium) gas near the center of a cluster. Many papers over the years have discussed the significant evidence supporting some version of this scenario
\citep{lf83,lee03,vanzee04a,vanzee04b,lis06a,lis06b,bos08,der10,kb12}
although properties of many dEs that cannot be explained by 
ram pressure alone (kinematics, shape, nuclei, metallicity)
has naturally led to questions on the role of 
ram pressure stripping in dwarf galaxy evolution 
\citep{fb94,con01}.
This clearly indicates that ram pressure stripping
is not the only mechanism important for producing many of the current dEs,
and that gravitational interactions must play a role, 
but the complexity has made it hard to clearly confirm the role of ram pressure stripping.

What has been lacking are clear examples of galaxies undergoing such a dI->dE transformation.
We propose that in the Virgo Cluster dwarf IC3418 we are witnessing
a critical stage in the transformation of a dI into a dE,
the recent and rapid removal of the entire ISM by ICM ram pressure stripping.



\section{The Galaxy IC3418}

\begin{figure*}[ht]
\includegraphics[height=0.45\textwidth,trim=0mm 0mm 0mm 0mm,clip]{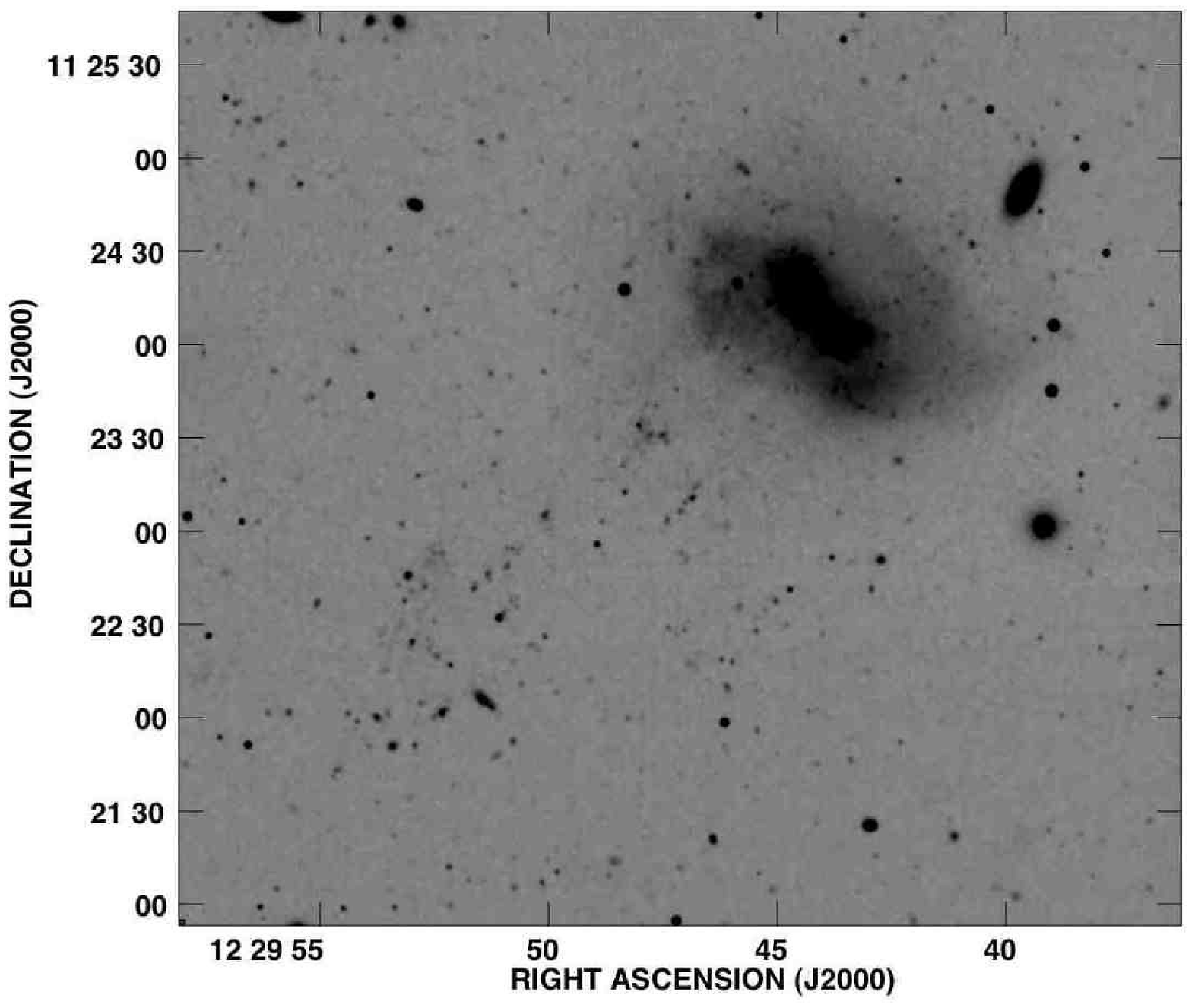}
\includegraphics[height=0.435\textwidth,trim=0mm 0mm 0mm 0mm,clip]{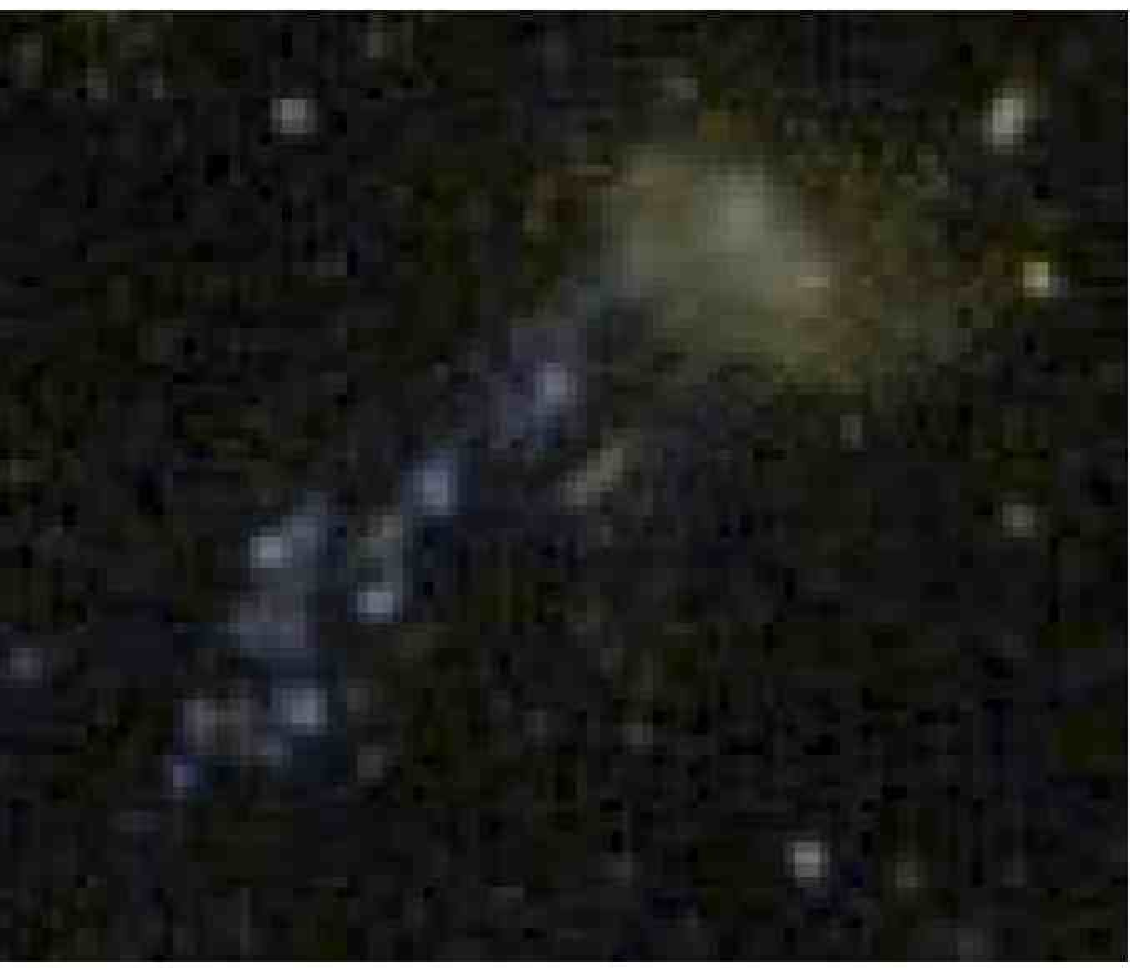}
\caption{
Comparison of IC3418 in optical and UV.
(right) WIYN B-band image of IC 3418, showing irregular
structure in main body of galaxy and numerous faint stellar associations
in a tail extending to SW.
(left) GALEX NUV+FUV color image of IC 3418, reproduced from \citet{hester10}.
FUV is blue, NUV is red, and the average UV intensity is green.
The bright UV emission from the tail indicates recent and ongoing star formation
from gas clouds which were presumably ram pressure stripped from the main body of the galaxy. Field of view in both images is $\sim5'x5'$.  The tail stands out from the background much more in the UV than the optical since the tail stars are young.
\label{fig1}
}
\end{figure*}

IC3418 (VCC1217) is a peculiar dwarf irregular galaxy (IBm?, \citep{rc3})
notable for an unusual UV-bright one-sided tail \citep{gil07,chung09,hester10,fuma11},
but highly deficient in HI \citep{hof89, chung09} and CO emission \citep{jac13}.
It is located close in projection (1.0$\deg$$\simeq$ 277 kpc$\simeq$0.17R$_{\rm vir}$ \citep{mclauglin99}) 
to M87 at the core of the Virgo cluster.

IC3418 has a line of sight velocity of 176$\pm$15 km/sec (\S 4.1.2),
consistent with a distance anywhere from the Local Group to the Virgo Cluster.
There is no current distance estimate for the galaxy, however 
the smoothness of the starlight in the main stellar body indicates that IC3418 is beyond the Local Group, 
since the pixel-to-pixel variation in brightness is much less in IC3418 than in Local Group galaxies. 
It is mostly on this basis, which is essentially the surface brightness fluctuation method, that 
\citet{bst85} considered IC3418 a true member of the Virgo cluster.
In this paper we adopt a distance of 16.7 Mpc, equal to the mean distance of galaxies in the M87 subcluster (cluster A) of Virgo \citep{mei07}.
Given its blueshift of 900 km/sec with respect to the mean Virgo velocity of 1079 km/sec (NED)
IC3418 is moving at high speed through the Virgo cluster toward us.
Its tail should therefore be behind the main body of the galaxy, and located further from us.

With a blue magnitude of B$_{\rm T}$$\simeq$14.5 and M$_B$$\simeq$-16.5, 
H-band luminosity = 8.7x10$^8$ L$\solar$ (from GOLDMINE, \citep{gav03}),
and a stellar mass of 4$\times$10$^8$ M$\solar$ \citep{fuma11},
it is among the brighter and more massive dwarfs in the cluster.
Its optical luminosity and stellar mass are about twice that of the SMC, and 
an order of magnitude less than NGC~4522, a Virgo cluster spiral galaxy 
with clear evidence for ongoing ram pressure stripping \citep{kvgv04}.

The optical and NUV-H colors (from GOLDMINE, \citep{gav03}) of the main body of the galaxy
place it within the blue cloud of galaxies \citep{hester10, fuma11} 
although its NUV-FUV colors are intermediate between the major "blue" and "red" galaxy populations
\citep{hester10,fuma11} , implying significant star formation until the last few hundred Myr.

In recent papers, \citet{hester10} and  \citet{fuma11} discussed UV and optical imaging of IC3418, 
in which they presented evidence that the galaxy is ram pressure stripped,
that the remarkable tail originates from star formation within ram pressure stripped gas, 
and that the galaxy is evolving from the blue cloud to the red sequence.
In this paper, we confirm and extend their main conclusions.
We present the first kinematic study of the tail features,
and provide important new information
on the morphologies and stellar populations of the tail features and the main body.
We emphasize that IC3418 is a clear example of a dwarf galaxy being transformed from 
a dI into a dE by ram pressure stripping,
and make the case that ram pressure stripping is the decisive process in converting dIs into dEs.

\section{Observations and Data Reduction}

 

\subsection{WIYN BVRHa Imaging}

\begin{figure*}[htpb]
\centering
\includegraphics[height=0.3\textwidth,trim=0mm 0mm 0mm 0mm,clip]{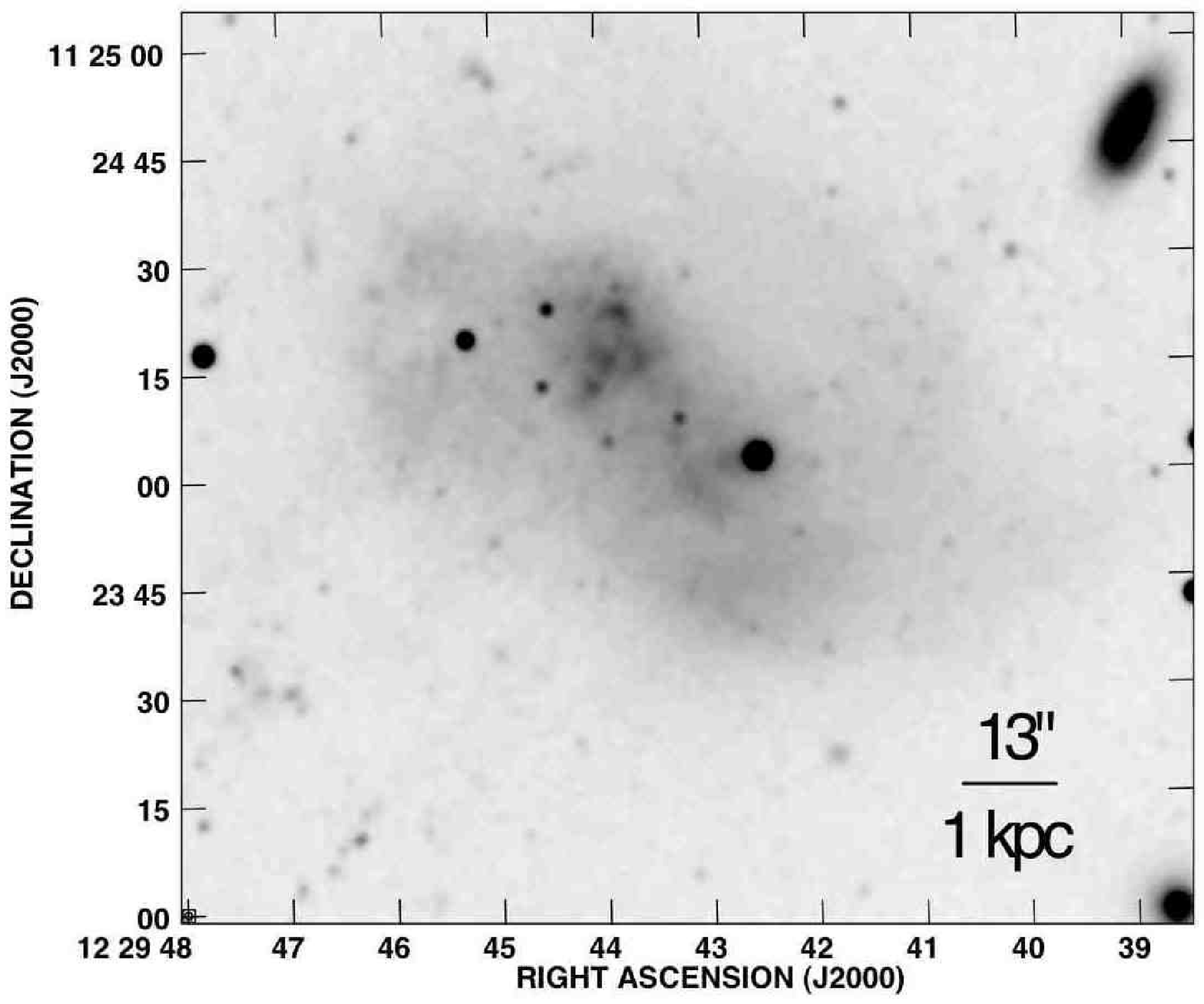} 
\includegraphics[height=0.3\textwidth,trim=0mm 0mm 0mm 0mm,clip]{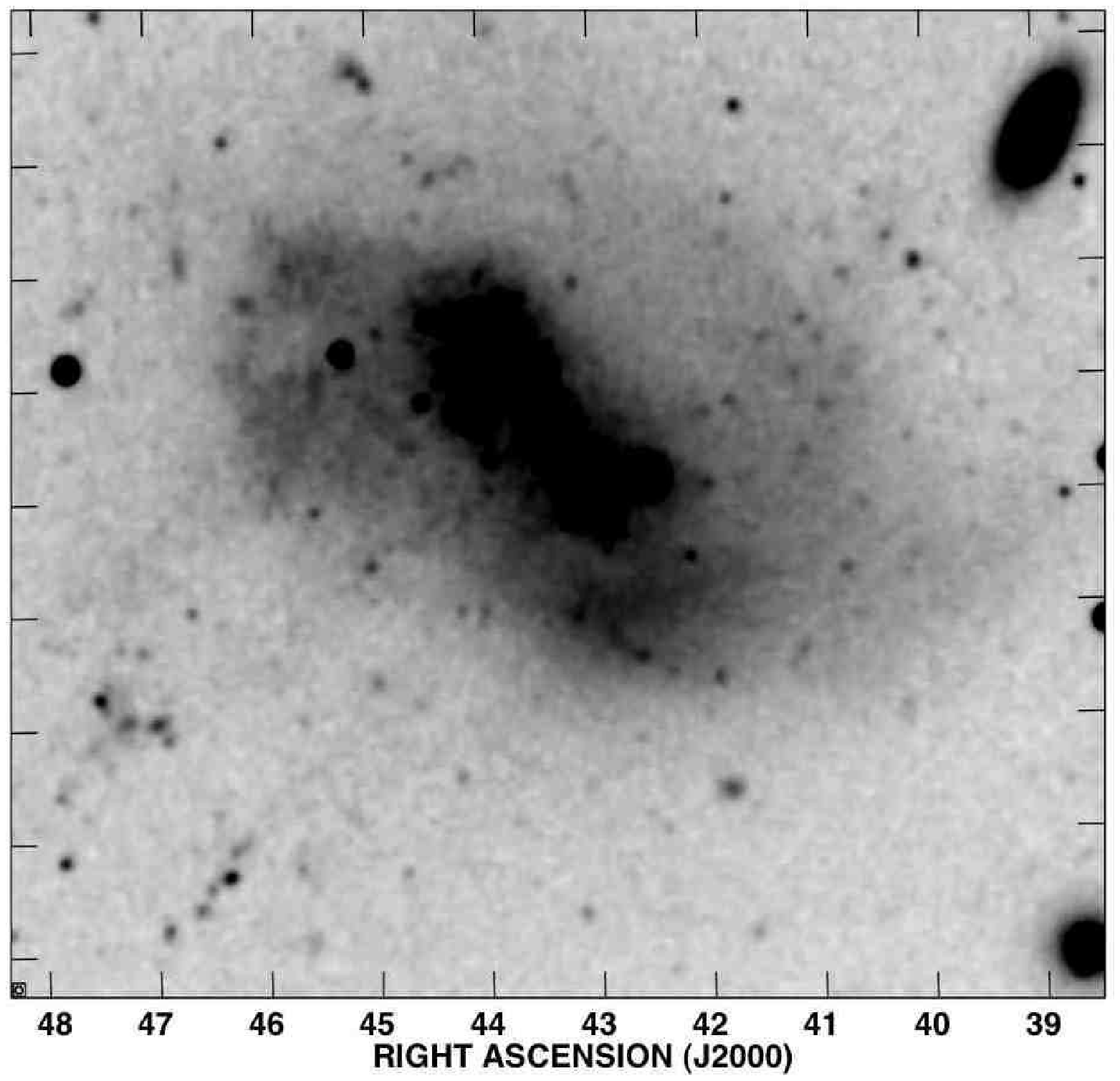} 
\includegraphics[height=0.3\textwidth,trim=0mm 0mm 0mm 0mm,clip]{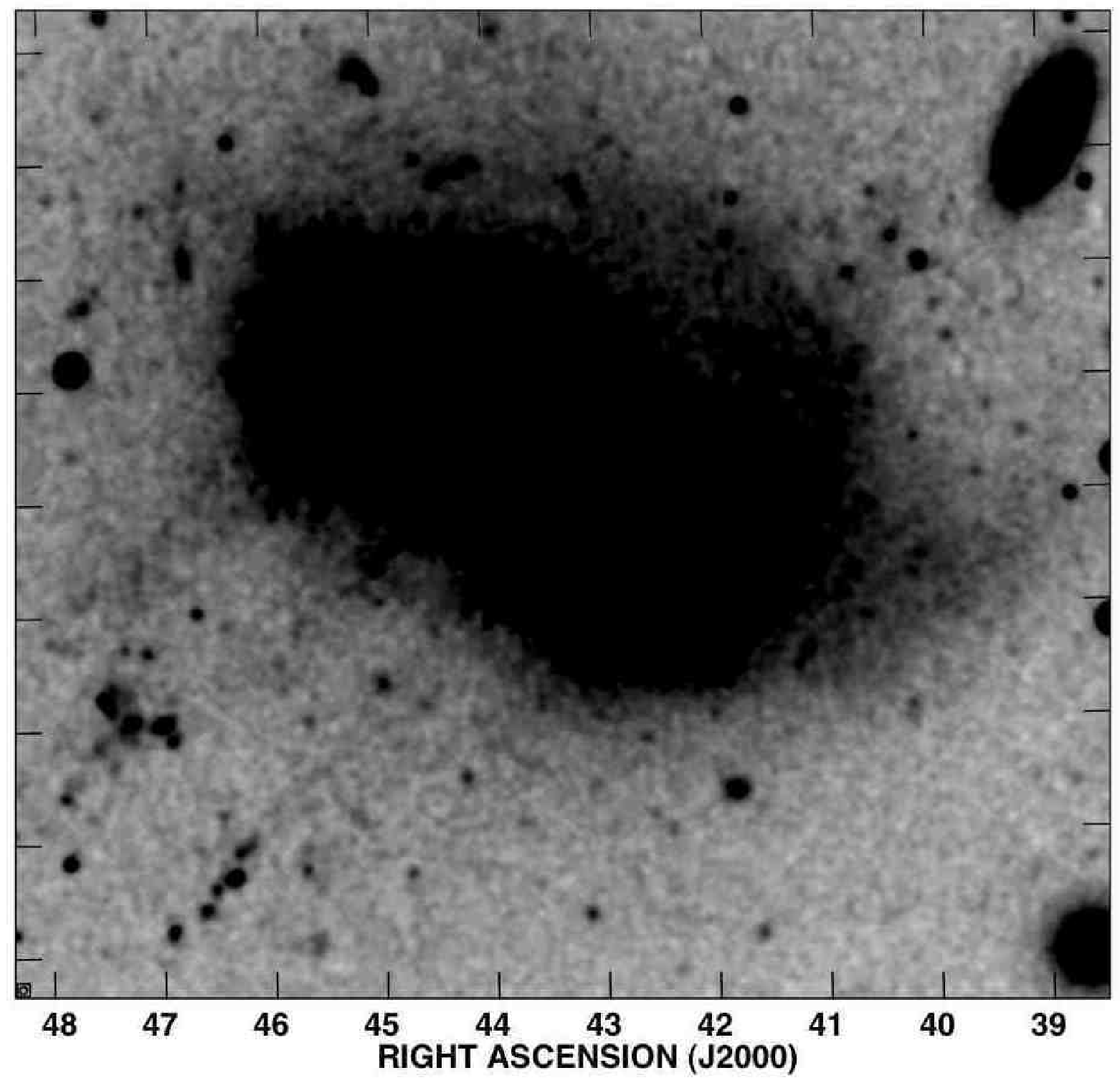} 
\caption{
WIYN B-band images of IC 3418, with different stretches to
highlight different regions of the main body of the galaxy.
a.) shallow image showing star formation substructure 
in central r=2 kpc. b.) medium image showing spiral arms and NW/SE brightness
asymmetry over r=2-4 kpc.
c.) deeper image showing spiral arm feature in outer disk plus the inner part of the tail.
\label{fig2}
}
\end{figure*}

Exposures of IC~3418 were taken in BVR and narrowband H$\alpha$+[N II]
filters on the 3.5m WIYN telescope$\footnote{
The WIYN Observatory is a joint facility of the University of Wisconsin-Madison,
Indiana University, Yale University, and the National Optical Astronomy
Observatories.}$
at KPNO in March 2009.
We used the 4096$\times$4096 pixel Mini-Mosaic imager, which has 2 CCDs,
and a plate scale of 0.14$''$/pixel,
giving a field of view of 9.6$'$ (42 kpc).
For the H$\alpha$ imaging,
we used a narrowband filter (W36) with a bandwidth of 60\AA\ centered on 6576\AA .
This filter includes the redshifted H$\alpha$ and the 2 straddling [N II] lines.

Integration times totalled 24 minutes for V and R, 
35 minutes for B, 56 minutes for the H$\alpha$+[N II] narrowband filter,
and were divided up into 5 or more dithered exposures per filter.
The seeing for all images used ranged between 0.8-1.0$''$.

 
We used the ``mscred'' package within IRAF to bias-subtract, flat-field,
register, and combine the images for each filter.
Flatfielding was done using both domeflats and 
skyflats produced from object images taken throughout the night.
Cosmic rays were removed using the pixel rejection routines in the
combine task. After sky subtraction, the R-band image was scaled and
then subtracted from the narrowband filter image
to obtain a continuum-free H$\alpha$+[N II] (hereafter referred to
as H$\alpha$) image.

 

\subsection{Keck Optical Spectroscopy: LRIS}

Optical spectroscopy of IC~3418 to study its stellar population was obtained with the LRIS
spectrograph on the Keck I telescope on 26-27 February 2009. The data
presented here use the blue 600/4000 grism to achieve wavelength
coverage of 3400-5500 \AA \, and spectral resolution of 6.1 \AA \, FWHM. Our
observations utilize a custom slitmask 1.5$\arcsec$ wide and 5.6$\arcmin$ long, with 2$\arcsec$ supports placed every 54$\arcsec$. The
seeing at start of the night was 0.8$\arcsec$. The slit was placed such that
it covered the entire optical extent of the galaxy and, on one side,
extended well beyond the galaxy to measure the sky emission. The slit
was oriented at 50 degrees, close to the major axis of the galaxy. Six
exposures each of length 1200 seconds were taken, for a total exposure
time of 7200 seconds.

The data were bias-corrected and flat-fielded using standard data
reductions techniques and IRAF packages. Following this, spatial
distortions in the 2D spectrum were removed using the IRAF package
TRANSFORM. This correction ensures that spectra could be reliably
extracted from the smooth light profile of the galaxy. These extracted
1D spectra were then wavelength corrected using Mercury-Cadmium-Zenon
arc lamps and had their sky signal subtracted using data from the end
of the slit, well beyond the optical extent of the galaxy.

\subsection{Keck Optical Spectroscopy: DEIMOS}

Optical spectroscopy of IC~3418 and its tail to study their kinematics
was obtained with the DEIMOS spectrograph  \citep{faber03a} on the Keck~II 10-m telescope on February 14, 2010.  One mask was observed with the 1200~line~mm$^{-1}$\,grating covering a wavelength region $6400-9100\mbox{\AA}$.  The spectral dispersion of this setup was $0.33\mbox{\AA}$ pixel$^{-1}$, equivalent to R=6000 for our $0$\arcsec$7$ wide slitlets, or a FWHM of $1.37\mbox{\AA}$. The spatial scale was $0$\arcsec$12$~per pixel.  Three exposures were taken, with a total exposure time of 3000 seconds.

Spectra were reduced using a modified version of the {\tt spec2d} software pipeline (version~1.1.4) developed by the DEEP2 team at the University of California-Berkeley for that survey. A detailed description of the reductions can be found in \citet{geha10}. The final one-dimensional spectra were rebinned into logarithmic wavelength bins with 15 $\kms$ per pixel.  Radial velocities were measured by cross-correlating the observed science spectra with a series of high signal-to-noise stellar templates. We applied a heliocentric correction to all velocities presented in this paper.

Our observations utilize a custom slitmask with a series of slits oriented at PA=135$\deg$, corresponding to the 
tail axis and approximately the minor axis of the galaxy.
One slit of 1.6$'$ length passed near the center of the galaxy, and includes the object S5 which is projected near the galaxy center.
Other shorter slits were centered on the brightest features in the tail, including most of those with H$\alpha$ emission detected in the imaging, plus some UV sources with little or no H$\alpha$ emission (U1, U3, U4). In addition we covered a couple of stars and background galaxies in the field.

The velocity of this galaxy has been uncertain. 
We measure the heliocentric velocity unambiguously 
from the absorption lines in the Keck DEIMOS spectra to be 176 $\pm$15 km/sec (\S 4.1.2).

\section {Results \label{results}}

\subsection{Main Body of IC3418}

\subsubsection{Imaging Analysis of Main Body}

A B-band image of the main body and tail of IC3418 is shown in Figure~\ref{fig1},
and B-band images of the main body are shown at different stretches in Figure~\ref{fig2}.
Overall the images show a low surface brightness galaxy
with substructure indicating recent star formation.

The left panel of Figure~\ref{fig2} shows a bright complex of arcs and
other features, $\sim$1 kpc in extent, located $\sim$1 kpc NE of the galaxy
center.
A second, fainter complex of similar size, exists $\sim$0.7 kpc SW of the center
(and E of the very bright star S6).
The central r$\simeq$2 kpc region is somewhat bluer in both the optical and the UV
than the surrounding regions \citep{hester10, fuma11},
indicating more recent star formation here.



The middle panel of Figure~\ref{fig2}
shows that at a surface brightness level which encompasses these 2 
young stellar complexes, the galaxy shows an elongated lima-bean shaped distribution 
resembling an irregular bar with length $\sim$2 kpc, which might be the origin of the
bar classification by the RC3.
There is no clear evidence for a stellar bar, and since the direction of elongation is 
perpendicular to the tail the elongation could be a result of ram pressure.
The disk from r=2-5 kpc (panels b and c) shows spiral structure,
further evidence that IC3418 is a disk system.
At these radii, the SE (tail) side of the main body is brighter and bluer
than the NW side,
perhaps due to recently formed stars in ram pressure stripped gas contributing 
additional blue light at the base of the tail.

Despite substructure in optical images resembling star forming regions,
the H$\alpha$ image shows no emission from the main body of the galaxy,
down to a 3$\sigma$ sensitivity limit of 2.1$\times$10$^{-17}$ erg s$^{-1}$ cm$^{-2}$ arcsec$^{-2}$,
which is 2.3 times deeper than \citet{fuma11}.
This corresponds to a point source H$\alpha$ luminosity of 7$\times$10$^{35}$ erg s$^{-1}$,
which is an order of magnitude fainter than the Orion Nebula.
HI emission \citep{hof89, chung09} is also undetected in the main body of the galaxy.
The only H$\alpha$ and HI emission detected are in the tail, which we discuss in \S 4.2.1 and  \S 4.2.2.
There is a marginal CO(2-1) detection suggesting a small amount ($\sim$10$^6$ M$\solar$) of molecular gas, 
but only from the central few arcseconds \citep{jac13}, and not from the young stellar complexes.
There is a suggestion of dust extinction in panels b and c of Figure~\ref{fig2}, on the SE side near the base of the tail,
where there is less light than at other regions at the same radii, although these surface brightness
variations might also be due to disk structure.
There are only upper limits on dust emission from IC3418 \citep{jac13}.

\begin{figure}[tb]
\epsscale{1.15}
\plotone{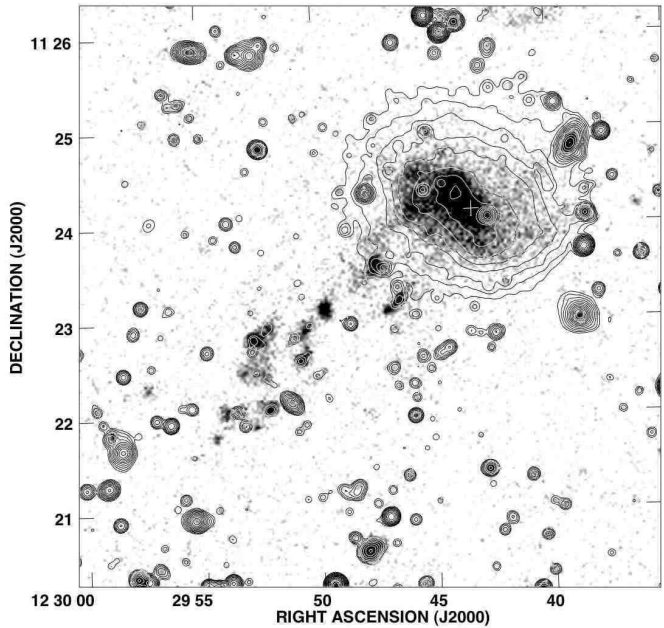}
\caption{
WIYN R-band image of IC 3418 convolved to 5$''$ resolution,
superposed on NUV image. The deep R-band image
shows a fairly regular stellar distribution at large radii
and no smooth stellar component to the tail, indicating that a tidal
interaction is not responsible for the tail. Outermost contour level is
26.5 mag arcsec$^{-2}$, and contour increment in 0.5 mag arcsec$^{-2}$. 
Location of Object S5, possibly a nuclear star cluster, is marked with a cross.
\label{fig3}
}
\end{figure}

In Figure~\ref{fig3} we show a ``deep'' optical image, our WIYN R-band image
convolved to 5$''$ resolution.
Residual flat-fielding errors may affect the outermost contour on the 
west side of the galaxy (right side of figure), but otherwise the
contour levels are robust.
This smoothed image reaches a surface brightness of 26.5 mag arcsec$^{-2}$.
At this level, shown by the outermost contours in  Figure~\ref{fig3},
the main body is seen to extend out to at least 80$''$= 6.2 kpc,
which is roughly twice the RC3 isophotal diameter.
The isophotes in the outer galaxy suggest a disk with PA$\simeq$50$\deg$ and 
inclination $\simeq$30$\deg$ .
The outer isophotes are fairly regular, with no evidence for tidal features.
The light from the main body encompasses $\sim$1/3 the tail,
however there are no suggestions of a smooth stellar component to the tail,
which would exist if the tail were a tidal tail.
A tail comprised of only gas and young stars, with no older stellar component,
is the sort of tail expected from ram pressure stripping.

Figure~\ref{fig3} shows that the light distribution in the inner galaxy is offset toward the SE (in the tail direction)
with respect to the outer isophotes.
This lopsided light distribution is consistent with ram pressure.

\newpage

\subsubsection{Stars \& Starlike Objects Across the Face of IC3418 and a Possible Nucleus}

\begin{figure}[htpb]
\epsscale{1.15}
\plotone{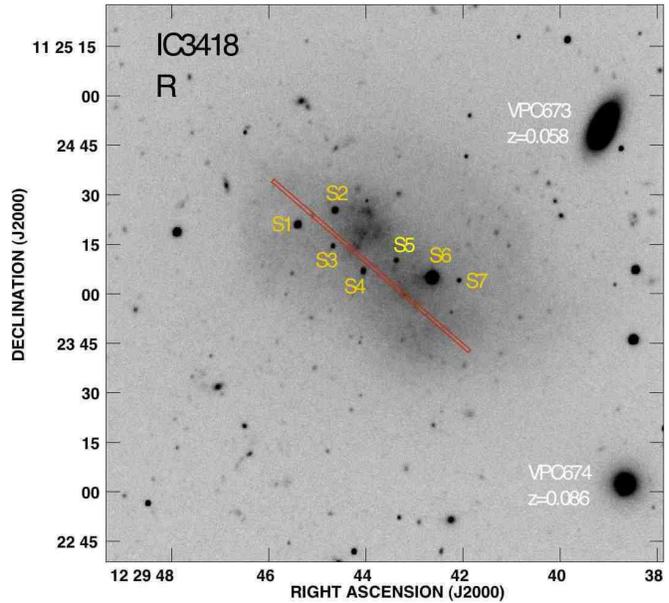}
\caption{
WIYN R-band image of main body of IC 3418, with bright stars and star-like objects
toward or in IC3418 identified. Object S5 has the velocity of IC3418, so it is in the galaxy. It is located within a few arcseconds of the center and could be the nucleus.
The other objects are either foreground stars, supergiants in IC3418, or globular clusters (see \S 4.1.2).
The position of the central $\pm$40$''$ of the Keck LRIS slit used for the stellar population analysis is marked.
\label{fig4}
}
\end{figure}

There are several bright stars or starlike objects across the face of the main body of IC3418,
concentrated within 30$''$ of the center of the galaxy.
We give the locations, magnitudes and colors of the 7 brightest ones in Table 1, and identify them on an R-band image in Figure~\ref{fig4}.
They are all spatially unresolved in our ground-based images, and for several of them it is not possible to tell whether they are foreground stars, supergiant stars in IC3418, or star clusters in IC3418.
At least one of these must be a foreground star since it is too bright to be a supergiant or globular cluster in IC3418 (S6).  Some are too bright to be supergiants in IC3418, but could be either foreground stars or star clusters in IC3418 (S1, S2). Others could be any of these possibilities (S3,S4,S7).


We have a spectrum of only one of them (S5), the object closest to the galaxy center.
The Keck DEIMOS spectrum shows a heliocentric velocity of 176$\pm$5 km/sec.
The galaxy light within the slit on either side of S5 has a velocity 181$\pm$5 km/sec,
clearly showing that it is located within the disk of IC3418.
We have measured velocities of a couple of Milky Way stars in the field near IC3418, and they have velocities of -4 km/sec and -25 km/sec, clearly different from IC3418.
S5 is unresolved spatially and spectrally, and with an absolute magnitude of M$_{\rm B}$=-9.35
it is consistent with either a supergiant, a globular cluster, or a nuclear star cluster \citep{geha02, cote06}. Its color of B-V=0.08 is quite blue. However the absence of Paschen lines in the spectrum suggests that it is not a blue supergiant, implying that it is likely a star cluster.

In early type galaxies, the nuclear star cluster is almost always located at the centroid of the inner light distribution  \citep{cote06}.
S5 is not at the centroid of the inner light distribution, but the inner light distribution in IC3418 is complex, not centrally peaked, and offset from the centroid of the outer light distribution.  It is close to, but offset by $\sim$5$''$ from the centroid of outer light distribution.
However given the disturbances to, and asymmetries in the galaxy, it could be the nucleus.
The absolute magnitude difference of 6.8 between S5 (M$_{\rm B}$=-9.35) and IC3418 (M$_{\rm B}$=-16.15, \citep{gav06}) is similar to that between nuclear star clusters and their host galaxies  \citep{cote06}.
Future studies which reveal the nature of this source are of 
interest because the origin of nuclear star clusters are not well understood  \citep{cote06}.



S5 is very close to the galaxy center and has a velocity close to that of the surrounding galaxy starlight.
We adopt its velocity as the velocity of the galaxy, although with a larger uncertainty to allow for a modest possible offset between S5 and the mean galaxy velocity.
Any offset is expected to be modest, since the rotation curves of dwarf galaxies rise very gradually.
Our adopted velocity is 176$\pm$15 km/sec.
This differs from the value of 38 km/s reported in the Goldmine database \citep{gav03}.


Some of the star-like objects could be supergiants (S3, S4, S7).
This is plausible given their high concentration near the galaxy center,
but we can't be sure as we do not have spectra of them, and they might also be foreground Milky Way stars
or globular clusters in IC3418.
Since supergiant stars have masses $\geq$10 M$\solar$ and lifetimes $\leq$100 Myr,
their presence would indicate some recent star formation in the central 2 kpc.
This could happen even with a mean star formation quenching age of 300 Myr (\S 4.1.3),
if star formation does not end uniformly in the galaxy, but some dense clouds survive the stripping of
surrounding material. This is seen to happen in some stripped spiral galaxies \citep{crowl05},
and allows a small amount of star formation to occur after most of the gas has been stripped.
This might occur especially in the galaxy center, the deepest part of the potential well.

\subsubsection{Stellar Population Analysis of Main Body}

\begin{figure}[htbp]
\includegraphics[scale=.37]{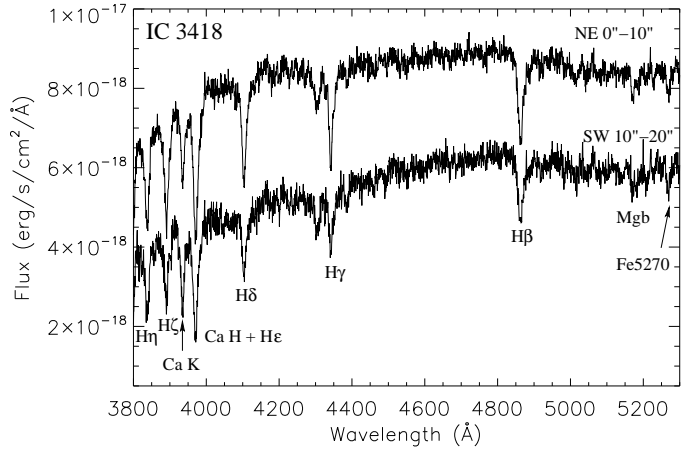}
\caption{
Keck LRIS spectrum in the blue for 2 regions in the main body of the galaxy.
No emission lines are detected, indicating a lack of gas.
The continuum is fairly blue and the Balmer lines are strong,
indicating that star formation was quenched 300$\pm$100 Myr ago,
and that a starburst occurred near the time of quenching.
The 2 spectra shown illustrate the spatial variations in stellar properties:
0-10$''$ in the NE has stronger lines and a bluer continuum than 10-20$''$ in the SW.
A constant offset of 2$\times$10$^{-18}$ erg s$^{-1}$ cm$^{-2}$ $\AA$$^{-1}$ has been added to the NE spectrum for display purposes.
\label{fig5}
}
\end{figure}

\begin{figure}[htbp]
\epsscale{1.2}
\plotone{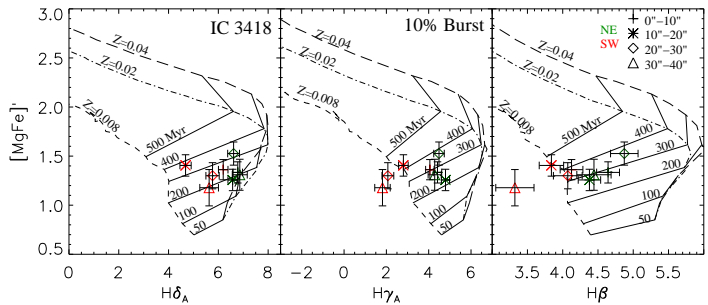}
\caption{
Lick indices and SB99 stellar population models for the main body of IC3418.
Model lines are shown for a star formation history with constant star formation until a quenching time, with a starburst at quenching time which formed 10$\%$ of the stars.
Dotted lines indicate stellar metallicities of Z=0.008, 0.02, and 0.04 (= 0.4, 1.0, and 2.0Z$\solar$).
Solid lines indicate time of quenching (measured from the present time). 
Symbols are shown for different radial segments of galaxy. 
The NE and SW sides have different line strengths, probably due to different burst strengths. 
A 10$\%$ burst is the best fit to the average of the 2 sides. 
\label{fig6}
}
\end{figure}

We use our Keck LRIS optical spectra together with GALEX UV colors \citep{martin05, gil07}
to analyze the stellar population of the main body of IC3418. Our Keck LRIS slit, with a width of 1$''$, passes through our adopted center along PA=50$\deg$. 
We analyze eight distinct 10$''$-long segments of the galaxy along the major axis,  extending from 0-40$''$ on both sides of center. The location of the slit is shown on a galaxy image in Figure~\ref{fig4}. 
Two sample spectra are shown in Figure~\ref{fig5}. 
The optical spectra show no obvious emission lines at any location, but strong Balmer stellar absorption lines, implying a lack of ionized gas, no ongoing star formation, and a post-burst stellar population, as found by \citet{fuma11}.

We measure line strengths using the Lick index definitions, which uses defined pseudo-continuum bands to avoid ambiguity in setting the continuum levels. 
In order to constrain the star formation history in the main body of IC3418, we compare the observed line strengths and UV colors to those predicted from  models generated from the Starburst99 (SB99) stellar population synthesis code \citep{leith99}.
For details see \citet{crowl08}.

We assume that there was constant star formation starting from a formation time (set to be 12 Gyr ago), followed by rapid quenching of star formation. We define the time since star formation ceased as the quenching time. In addition, we have considered the effect of adding a starburst at the time of quenching, in which 0-20$\%$ of the stellar mass is formed in a burst at the time of quenching (equivalent to an increase in the star-formation rate (SFR) of a factor of 1-50 for a burst duration 50 Myr). The strong Balmer lines clearly indicate a significant recent starburst. A 0$\%$ burst strength is inconsistent with all Lick indices, and the strong H$\delta$ lines indicate a burst strength of at least 5$\%$.

\begin{figure}[htbp]
\epsscale{1.2}
\plotone{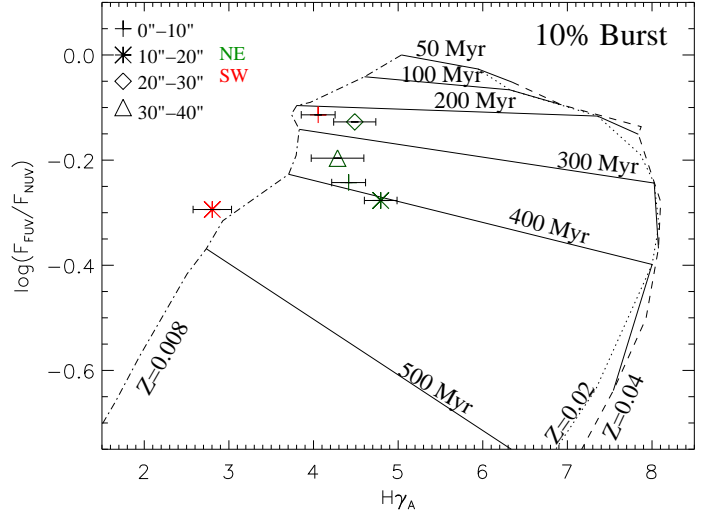}
\caption{
GALEX FUV/NUV flux ratios versus the H$\gamma$ Lick index
for the main body of IC3418.
Model lines are shown for a star formation history with constant star formation until a quenching time, with a starburst at quenching time which formed 10$\%$ of the stars.
Dotted lines indicate stellar metallicities of Z=0.008, 0.02, and 0.04 (= 0.4, 1.0, and 2.0Z$\solar$).
Solid lines indicate time of quenching (measured from the present time). 
Symbols are shown for different radial segments of galaxy. 
The UV flux in the outer 2 radial bins on the SW side was too faint to accurately measure in our small apertures.
\label{fig7}
}
\end{figure}

Figure~\ref{fig6} shows the observed strengths of several Balmer Hydrogen lines versus the metal [MgFe]' Lick index as a function of position.
The lines in Figure~\ref{fig6} indicate model line strengths for various quenching times and metallicities, assuming a burst strength of 10$\%$, which gives the best overall fit to the galaxy. 
Figure~\ref{fig7} shows the FUV-NUV colors, 
measured from the GALEX images \citep{martin05, gil07}
in apertures close in size to those used for the Lick indices,
versus the strength of the H$\gamma$ line, with lines indicating the same models. Since the Lick indices and the FUV/NUV ratio have different dependencies on the burst strength, quenching time, and metallicity, using both gives more robust constraints on these parameters and also gives a measure of the uncertainties.

At a given Balmer line strength, the metal line index [MgFe]' is much weaker than would be expected for solar metallicity, thus the stars clearly have a sub solar metallicity. 
Taking the average values for the 3 Balmer lines\footnote{
H$\beta$ is sometimes a less reliable indicator than the higher order Balmer lines, since the stellar absorption line is more likely to be partially filled in with residual emission from gas. Since none of the spectra show evidence for emission lines in the main body of the galaxy, we think this should be a minor effect in IC3418, so we use all 3 Balmer lines. }
the Lick index plots (Figure~\ref{fig6}) show a stellar metallicity of Z=0.012=0.6$\pm$0.2 Z$\solar$, 
with no clear evidence for a metallicity gradient.

We find a star formation quenching time of 300$\pm$100 Myr, taking the average of the Lick indices for the 3 Balmer lines and the FUV/NUV ratio versus H$\gamma$. This is consistent with the 200$\pm$90 Myr estimated by \citet{fuma11} from the FUV/NUV ratio, although a bit older than their value perhaps because the quenching time estimate depends on both the metallicity and burst strength, which we have estimated quantitatively.

An older quenching time at larger radius is expected for ram pressure stripping, which operates from the outside in.
The azimuthally averaged UV colors given in \citet{hester10} show a clear but modest outward gradient from FUV-NUV=0.65 at r=6$''$ to FUV-NUV=1.00 at r=48$''$. This would correspond a gradient in quenching time of $\leq$70 Myr from 0-48$''$.
However neither the Lick indices in Figure~\ref{fig6} nor the UV ratios in Figure~\ref{fig7} show a clear radial gradient in quenching time.
Our apertures don't show a uniform outward gradient in UV colors,
and several of them have bluer colors than any of the azimuthally averaged values.
This probably reflects the fact that our apertures cover 
localized regions with more active recent star formation.
While there many be an average radial gradient in quenching time, 
the quenching time may not vary smoothly with position since
ISM is not smooth, and localized regions of higher density can resist stripping longer than 
lower density regions closer to the center.
These localized spatial variations in star formation histories 
may mask any clear radial gradient in quenching time.
We adopt an upper limit on the quenching gradient of $\leq$70 Myr from 0-48$''$. This is a modest gradient which implies that the galaxy was stripped quickly. In a simple model of stripping in IC3418, it takes $\sim$150 Myr to strip gas with $\sigma$=15 M$\solar$ pc$^{-2}$ from r=3 to 0 kpc \citep{jac13}. The stripping timescale in IC3418 is similar to, and perhaps quicker than this.

Much larger than any radial gradients in IC3418 are spatial differences in optical-UV colors and line strengths 
between the NE and SW sides.
The Lick index values on SW side of the major axis are offset to lower Balmer line strengths than the NE side. Correspondingly, the optical images and g-i color map shown by \citet{fuma11} show that the NE part of the galaxy is significantly bluer than the SW part.  
The shift in the Lick index plots is inconsistent with a difference in quenching time,
and consistent with a difference in burst strength or metallicity.
We regard as a difference in metallicity between the 2 sides as unlikely, and think the difference is more likely attributable to a difference in burst strength.
A burst strength of $\sim$7\% of the SW side and $\sim$13\% on the NE side would account for the difference.
On small scales, variations in quenching time and burst strength are to be expected, especially in dwarf galaxies.
It is unclear whether these spatial differences are related to the stripping event, as such variations might also be expected in undisturbed dwarf galaxies.
A burst strength of 10\% provides a good fit to the Lick indices averaged over all apertures,
although the true strength of the burst could be somewhat higher or lower than 10$\%$, as this
depends on the true star formation history in IC3418, which is likely not as simple and smooth as our assumed
star formation history.

\begin{figure}[htbp]
\epsscale{1.2}
\plotone{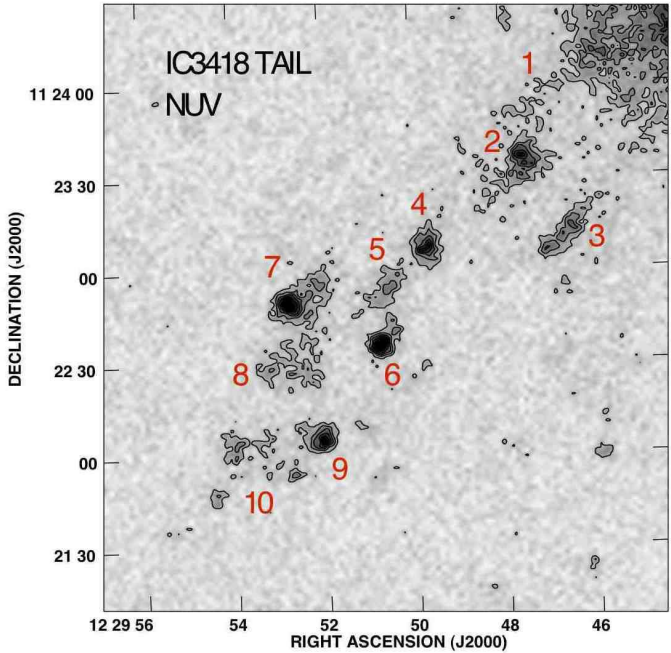}
\caption{
GALEX NUV greyscale and contour image of tail of IC3418, with distinct
regions numbered.  Eight of the ten UV regions are elongated with position angles
similar to the large scale tail. 
\label{fig8}
}
\end{figure}

In order to test the sensitivity of the quenching time and burst
strength to the assumed star formation history, we have run models
with alternate star formation histories.
One model adopts the mean star formation history of dwarf galaxies in
the ANGST sample (Fig. 11 of \citet{weisz11}), which is well approximated
by an exponentially declining function with a timescale of 1.25 Gyr,
plus a constant term, scaled such that 65\% of the stellar mass is
formed by z=0.7 (6.5 Gyr ago).  With such a baseline star formation
history, we find fitting the Lick indices and FUV/NUV ratios in IC3418
require a very similar quenching time of 250 Myr and the same burst
strength of 10\%.
Since there is significant variation in the star formation histories
of individual dwarf galaxies, and many have experienced significant
bursts during their history, we have also run models with a baseline
star formation history given by the lower envelope of the ANGST dIs
shown in Fig 6 of \citet{weisz11}, plus a burst at intermediate times, in
addition to the recent burst at the time of truncation.  In this case,
fitting the Lick indices and FUV/NUV ratios in IC3418 requires a very
similar quenching time of 200 Myr and a 5\% burst strength at
quenching time.  The precise timing and strength of any burst at
times earlier than 1 Gyr ago doesn't
significantly affect the Lick indices or FUV/NUV ratios. 
The baseline
star formation history in this case has a relatively low star
formation rate over most of the galaxy's history, and a large amount
of star formation ($\sim$30\% of the total stellar mass) in the last 1 Gyr.
Thus the only way to have a burst strength much lower than $\sim$10\% at
quenching time is to have another large recent burst in the last $\sim$1
Gyr. Even in this extreme case, the burst required at quenching time
is $\sim$5\%. 
In summary, the precise timing and strength of the recent starburst
depend only a little on the assumed star formation history before $\sim$1 Gyr ago, and even with these alternate star formation histories, the quenching time and burst strength fit within the adopted values of
300$\pm$100 Myr and 10$\pm$5$\%$.

Other ram pressure stripped galaxies are observed to experience, or have experienced, starbursts at the time of stripping \citep{ck06,abram11,mer13}. A spiral galaxy experiencing ram pressure stripping in the rich cluster A3558 has an ongoing starburst over part of the disk with a 5 times increase in the SFR over the past 100 Myr, plausibly triggered by ram pressure \citep{mer13}.
Among the Virgo spiral galaxies studied spectroscopically by  \citet{crowl08}, the one with spectral properties most similar to IC3418 is NGC 4522, a spiral with clear evidence for ongoing ram pressure stripping. The gas-stripped outer disk of NGC 4522 has the strongest Balmer lines of all the Virgo spirals studied by  \citep{crowl08}, as it was stripped most recently (100 Myr ago) and had a starburst at the time of truncation. The NGC 4522 starburst formed 2$\%$ of the stellar mass, corresponding to a 5 times increase in the SFR over 100 Myr, the same intensity as the starburst in the A3558 spiral  \citep{ck06}. IC3418 has even stronger Balmer lines than NGC4522, but redder FUV-NUV colors, suggesting an older quenching time but a stronger burst, with an amplitude of $\sim$10$\%$ of the stellar mass.
This may be because star formation is generally burstier in dwarfs than spirals, but might also be due a more extreme ram pressure event in this fully-stripped dwarf.

\newpage

\subsection{Tail of IC3418}

\subsubsection{Tail Morphology in UV, Optical and H$\alpha$}

GALEX UV images of IC3418 \citep{chung09,hester10,fuma11} show a remarkable one-sided tail of young stars
extending 17 kpc outward from the main body of the galaxy (Figure~\ref{fig1}).
At the GALEX resolution of 6$"$=0.46 kpc,
there are $\sim$10 distinct bright UV sources within the tail, with morphologies of diffuse regions, knots, 
linear streams, or head-tail sources, which are a combination of knots and linear streams.
While \citet{hester10} and \citet{fuma11} have previously described the tail,
we point out several key features of the tail morphology not described by them.
In  Figure~\ref{fig8} we label 10 UV tail features, numbered sequentially with increasing distance from galaxy.
Our labeling is different from and simpler than both \citep{hester10} and \citep{fuma11}.
We don't create separate lists for "knots" and "filaments" or "diffuse regions", since there is overlap and some sources show characteristics of more than one morphology.
For example, UV7 has a head-tail morphology (discussed below). \citet{fuma11} label the head as K5 and the tail as F2, whereas we label it as one source, since we think that better reflects its nature.


The 7 brightest UV tail features are elongated and nearly parallel to each other, 
with position angles 125-138$\deg$, which matches the PA of the large-scale tail, 132$\pm$1$\deg$.
The longest stellar stream is ~25$''\simeq$2 kpc in length.
An 8th feature, UV1, the UV tail stream closest to the galaxy, 
has a PA offset by $\simeq$10$\deg$ from the others.
This feature, which was not identified by \citet{hester10} or \citet{fuma11},
is faint and has redder UV colors, it appears slightly curved, and may be an ``older'' 
stellar stream now falling back into galaxy.


WIYN images show optical counterparts for all the UV sources,
although as seen in Figure~\ref{fig1}, the tail sources stand out from
the background galaxies much more in the UV than the optical,
indicating recent star formation.
The structure of the stellar streams contain information
on the numbers and sizes and motions 
of dense clumps of star-forming gas, and 
the length of stellar streams contain information on 
how long dense gas lumps retain an identity.
Within the kpc-scale NUV sources, the higher resolution optical images reveal
complexes of optical knots with characteristic lengths of 0.5-3 kpc
and widths of 200-400 pc, rather than simple monolithic streams.
The optical substructure suggests
multiple centers of star formation within outwardly accelerating 
200-400 pc diameter gas concentrations.





\begin{figure}[htbp]
\plotone{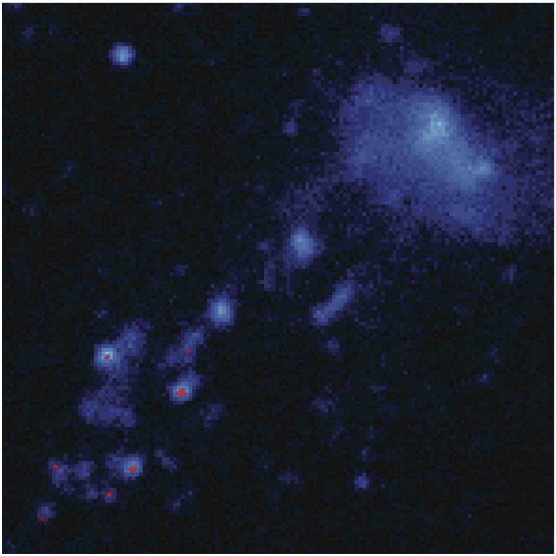}
\caption{
H$\alpha$+[NII] (red) plus NUV (blue) image of IC3418,
showing that all bright HII regions are in the outer part of the tail.
\label{fig9}
}
\end{figure}

\begin{figure}[htbp]
\epsscale{1.2}
\plotone{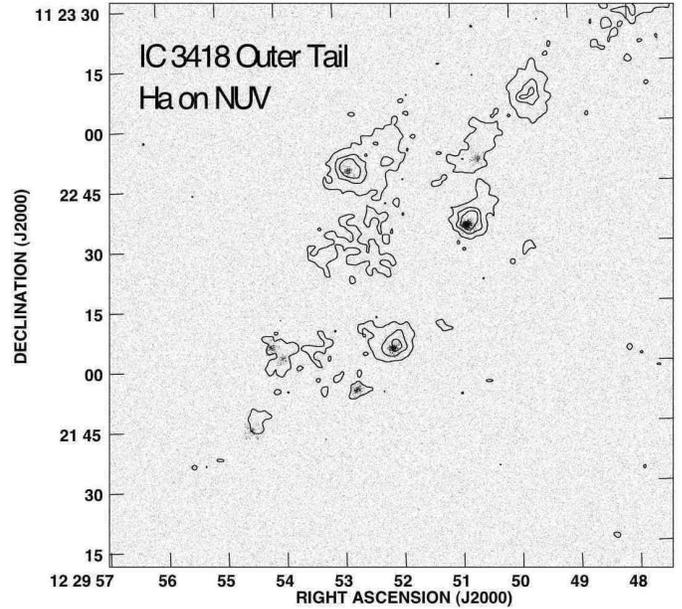}
\caption{
H$\alpha$+[NII] line image on NUV contours of outer tail,
showing HII regions. Note bright HII regions at heads of
head-tail UV sources, and the outward offset of the HII regions relative to the UV peaks.
\label{fig10}
}
\end{figure}

The only H$\alpha$ emission detected in the imaging is from discrete HII regions
in the outer half of the tail, at distances ranging from 10-17 kpc from the galaxy center.
These regions are faint, and at this level the continuum-subtracted image
also shows artifacts from bright continuum sources.
We have individually examined all features from the continuum-subtracted image, 
in both the broad and narrowband filter images to learn which features
are real H$\alpha$ sources. Based on the ratio of counts in the
broad and narrowband filter images, we identify only 8 ``true'' HII regions,
which are the same ones shown by \citet{fuma11}.
We masked out everything but the confirmed HII regions to make the H$\alpha$-NUV overlays in Figure~\ref{fig9} and Figure~\ref{fig10}.
Fluxes of all HII regions are given in Table 2.
The total H$\alpha$+[N II] flux of all the tail HII regions is 3.0$\times$10$^{-15}$ erg s$^{-1}$ cm$^{-2}$.
This total flux is consistent with \citet{fuma11},
although we find significantly different fluxes for some of the individual HII regions.

Spectroscopy has revealed 3 faint emission line sources not detected in the imaging,
including 2 in the inner tail, located within UV3 at $\sim$6 kpc from the galaxy center, and one in UV9, a few arcseconds closer to the galaxy than the bright HII region detected in UV9.
These are approximately an order of magnitude fainter than the faintest HII regions detected in the imaging,
and are likely associated with older and less massive stars. They have H$\alpha$ luminosities of $\sim$10$^{35}$ erg s$^{-1}$, and at least the two brighter ones (which have more than 2 lines detected) have line ratios similar to those of HII regions in the outer tail, with H$\alpha$$>>$[NII] (see Figure~\ref{fig13}).  With such luminosities and line ratios, these emission line sources could be classical HII regions, powered by single B stars photoionizing the surrounding gas remaining from the birth cloud, or single Be stars, which photoionize gas ejected from the star, or planetary nebulae. B2-B4 stars have main sequence lifetimes of 30-100 Myr, and the stars producing planetary nebulae are generally even older. Thus the low luminosity emission line sources in the inner tail are consistent with recent but not ongoing star formation. 

The overlays of the H$\alpha$ and NUV images in Figure~\ref{fig9} and Figure~\ref{fig10}
shows that all H$\alpha$ sources are coincident with NUV sources.
Three of the brightest HII regions are located at the heads of UV head-tail sources,
with tails pointing back toward the galaxy (UV6, UV7, and UV9).
Figure~\ref{fig10} shows that the H$\alpha$ and UV peaks are offset, with the
H$\alpha$ located 1-2$''$=80-150pc further from the galaxy than the UV peak.
UV6 and UV7 exhibit clear head-tail morphologies, and 
in UV9, the outermost bright Ha/NUV source, the NUV "tail" is very short
but there is a clear  offset between the H$\alpha$ and NUV peaks,
in the same direction as the tails.
This is evidence for gas preferentially accelerated and dynamically decoupled from
newly formed stars \citep{hester10}.

\begin{figure}[htbp]
\epsscale{1.2}
\plotone{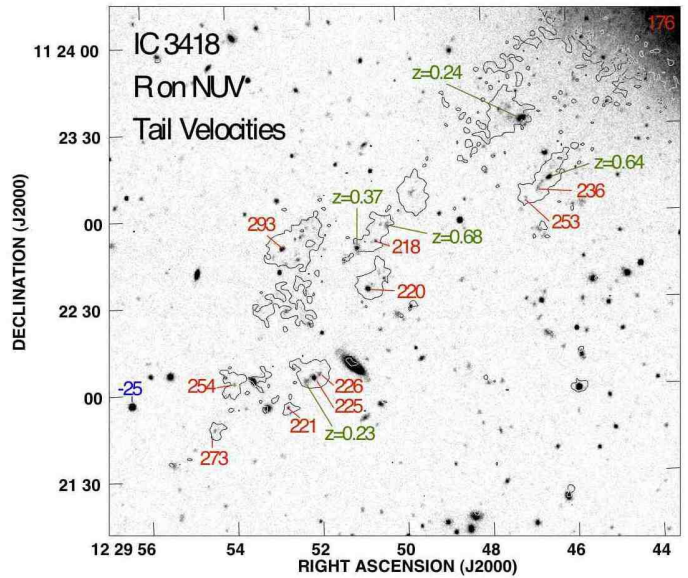}
\caption{
Velocities of sources in the tail region of IC3418 plotted on R-band greyscale image with contours of NUV emission. Red indicates heliocentric velocities (in km/sec) of emission line sources (HII regions) in the tail of IC3418, blue indicates velocities of foreground stars, green indicates redshifts of background galaxies.
\label{fig11}
}
\end{figure}

Correcting for the presence of contaminating background galaxies 
shows that possibly 4 of the outer tail HII regions are located near the heads of UV streams, with an outward offset
(UV5 as well as UV6, UV7, and UV9).
The innermost HII region in the outer tail is located 130$''$ = 10 kpc from the nucleus,
near the middle of a linear NUV feature (UV5). However Figure~\ref{fig11} shows that background galaxies
form the outermost and innermost parts of UV5. Removing them would likely put the HII region in UV5 near the head 
of a stellar stream. Similarly, the outermost part of UV9 coincides with a background galaxy, so 
the portion of UV9 that is associated with IC3418 has an even clearer head-tail morphology than suggested by Figure~\ref{fig10}.

Not all of the HII regions are at the heads of head-tail features.
Four HII regions are in the outermost part of tail, 
beyond the brightest UV sources, in a triangle-shaped feature 
of lower surface brightness UV emission (UV10).
The outermost  H$\alpha$ and UV source detected is at the 
tip of this triangle, 218$''$=17 kpc from the nucleus.
These outermost 4 HII regions are faint in both H$\alpha$ and NUV,
so the outer tail has modest active star formation and 
not much stellar mass formed.

Of the 6 brightest NUV sources in the tail,
the outer 3 all have H$\alpha$, and the inner 3 have little or no H$\alpha$.
These observations suggests 3 distinct zones in the tail, 
with different H$\alpha$/UV ratios:
the inner tail (inner 2$'$, UV1-4) has bright UV but little or no 
H$\alpha$,
with an H$\alpha$/NUV ratio $<$10$\%$ that of the outer tail,
the outer tail (2-3$'$, UV5-9) has bright UV and modest H$\alpha$,
and the very outer tail (3-4$'$, UV10) has fainter UV and modest H$\alpha$,
with an H$\alpha$/NUV ratio 1.6 times that of the outer tail.
The inner tail UV sources show no obvious systematic difference
from the outer (middle) tail UV sources in UV or optical color,
whereas the very outer tail features have bluer colors \citep{fuma11}.

The morphologies of the tail features differ in the 3 zones.
The inner tail has UV-bright linear streams and knots, but no head-tail features and very little H$\alpha$ emission.
The outer tail has UV-bright head-tail features, i.e., knots with H$\alpha$ at the heads of linear stellar streams ("fireballs").
The very outer tail has lower surface brightness diffuse UV features with associated H$\alpha$ emission.
We propose the following interpretation for the different tail zones.
Nearly all the gas has been stripped from the inner tail, and some stellar features here may be falling back into the galaxy.
The outer tail still has gas and ongoing star formation, with some large gas concentrations being outwardly accelerated, forming fireballs.
The very outer tail still has gas and ongoing star formation, but smaller gas concentrations, so less bright stellar knots.
Part of this is likely due to the spatial segregation of gas clumps of varying densities -- 
at a given ram pressure, lower density clumps are accelerated more and get further from the galaxy \citep{tb10,tb12,jac13}.
This could help explain the difference between the outer and very outer tail.

Some features of the UV-H$\alpha$ morphology in IC3418 are like those in the ``Fireball Galaxy'' RB199 in the Coma cluster, in which UV-bright stellar streams in the outer part 
of its tail have associated H$\alpha$ emission,
but those closer to the galaxy center are seen only in the UV \citep{yos08}.

The width of the tail in IC3418 is relatively constant with radius. Simulations without radiative cooling show flaring of the tail \citep{rb08}, whereas those with cooling show narrower tails with nearly constant width \citep{tb10}. The presence of star formation in the tail is a clear signpost of cooling, and thus is likely related to the narrow width of the tail. The tail width of 45$''$=3.5 kpc is similar to the lateral extent of the region in the main body of the galaxy showing evidence of recent star formation. This may explain the tail width -- it has roughly the same lateral extent as the part of the galaxy that was most recently gas-stripped.

Deeper observations should show evidence of tail/halo stars further out from
the centroid of the tail than the bright UV features, reflecting stars formed
from gas stripped from the galaxy's outer disk. Such stars would have fallen
back more toward the galaxy. Stream UV1 may be such an example --
its stars are older, the stream is furthest from the tail midline, and it is the
most curved stellar stream.
It could be a stream formed from gas further from the galaxy center, stripped at
an earlier time.

The H$\alpha$ luminosity of the HII regions in the tail
corresponds to a star formation rate
of 1.9$\times$10$^{-3}$ M$\solar$ yr$^{-1}$.
This is a small amount.
\citet{fuma11} estimated the mass of newly formed stars in the tail
to be $\sim$2.5$\times$10$^6$ M$\solar$ over $\sim$200 Myr,
from fits to the UV-optical SEDs.
This is less than 1$\%$ of the galaxy's stellar mass,
and suggests that stars formed in ram pressure stripped tails do not
make a large contribution to the intracluster stellar population in Virgo.
Indeed, most of the gas stripped from Virgo spirals
remains gas and joins the ICM \citep{vollmer12}.

\subsubsection{HI in the Tail of IC3418}

\begin{figure}[htbp]
\centering
\includegraphics[height=0.4\textwidth,trim=0mm 0mm 0mm 0mm,clip]{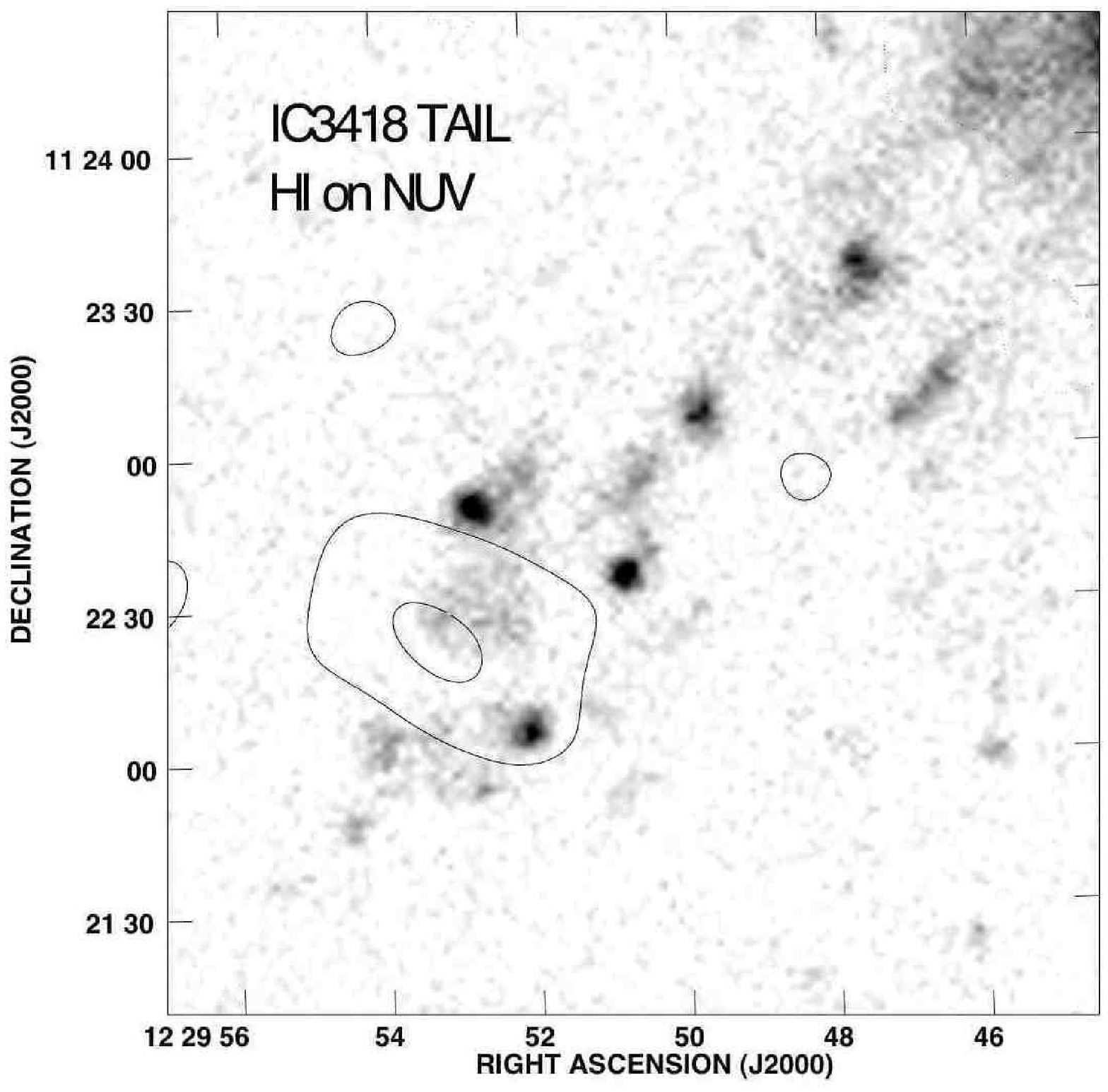} 
\includegraphics[height=0.4\textwidth,trim=0mm 0mm 0mm 0mm,clip]{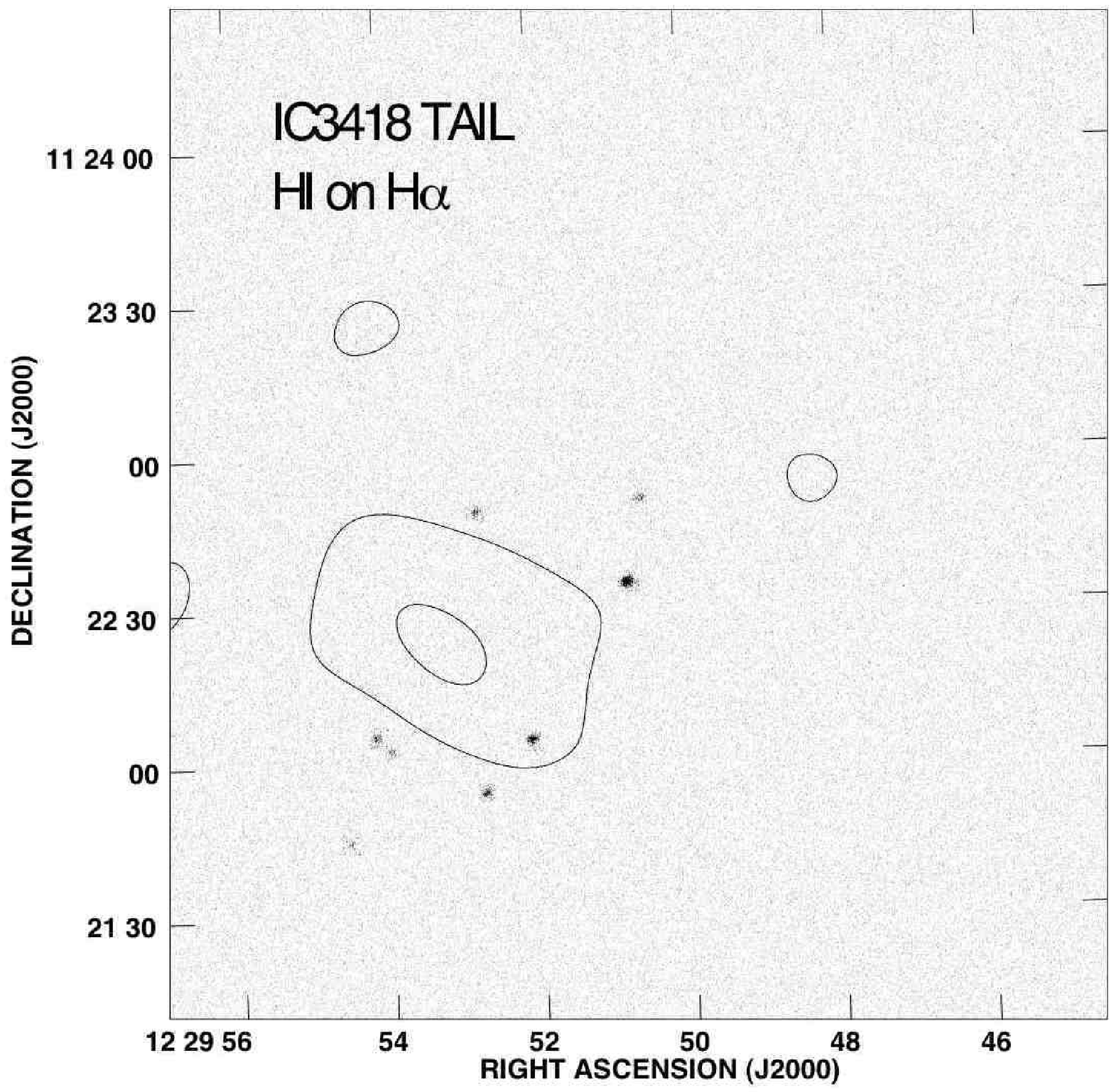} 
\caption{
Map of HI channel centered at 234 km/s with width of 10 km/s overlain on 
a.) NUV image, and 
b.) H$\alpha$ image of IC3418.
Contour levels are -2, 2, 4$\sigma$, where 1$\sigma$=0.6 mJy/beam = 0.06 M$\solar$ pc$^{-2}$ = 8$\times$10$^{18}$ cm$^{-2}$.
\label{fig12}
}
\end{figure}

We have detected weak HI emission in the outer tail from the VIVA data \citep{chung09}.
While we originally reported no HI emission in IC3418  \citep{chung09}, we reinspected the VIVA data cube and carefully examined those portions of the cube which matched the positions and velocities of the HII regions in the tail.  
When we spatially smooth the cube to a resolution of 30$''$, we find emission in one 10 km/s width channel with a peak intensity of 4$\sigma$,  where the rms is 0.6 mJy/beam.
This weak but statistically significant HI feature peaks at a velocity of 234 km/s, 
and has an HI flux of 0.038$\pm$0.007 Jy km/s.

An overlay in  Figure~\ref{fig12} of the HI channel at 234 km/s
on the NUV and H$\alpha$ images show that the HI emission is 
located in the outer part of the tail and is extended in the direction perpendicular to the tail.
The HI peak is centered closest to diffuse region UV8,
and covers the UV/H$\alpha$ head-tail source, UV9.
It is located just beyond the the bright UV/H$\alpha$ head-tail sources, UV6 and UV7,
and lies just inside the outermost source UV10 and its HII regions.
While many of the tail HII regions are slightly beyond the region where HI has been detected,
it is likely that the detected HI is merely the peak of a more extended HI distribution which covers the HII regions.
The HI velocity of 234$\pm$5  km/s is generally consistent with the HII regions in the tail,
and is very close to that of UV9, which has v=225-226 km/s (as discussed below).

The detected flux corresponds to 3.7$\times$10$^7$ M$\solar$, which is 
only $\sim$6$\%$ of the expected HI mass of a typical dI with the mass of IC3418 \citep{gav05}.
Most of the stripped gas must be more spread out, or in another phase.

Our reinspection of the HI data cube, involving smoothing with different combinations of spatial and velocity averaging, still shows no HI in the main body of galaxy, 
and we place an upper limit similar to that reported in \citet{chung09},
of 6$\times$10$^6$ M$\solar$ of HI (3$\sigma$) assuming a 100 km/s profile 
per 20 arcsec beam.
The velocity range of the galaxy is now better known than at the time of the original VIVA analysis,
when the reported velocity of 38 km/s from the GOLDMINE database \citep{gav03}
was very close to Galactic HI emission.
The velocity range of the galaxy and tail is now known to be 140-290 km/s, outside the range of Galactic HI.
Thus our upper limit is now more reliable as the VIVA HI data cube has
good data over the entire velocity range of IC 3418.
With this upper limit, the main body of IC3418 has less than 1$\%$ of its expected HI mass.

\subsubsection{Kinematics of Tail Sources in IC3418}

With the Keck DEIMOS spectroscopy, we have detected emission lines from every targeted HII region seen in the imaging, plus weak emission lines in the UV feature (UV3) near the galaxy which was undetected in the H$\alpha$ imaging. We have also detected emission + absorption line spectra from several background galaxies in the tail region.  The brightest optical sources within U2, U3 and U5 are all background galaxies with z=0.23-0.68. We discuss in \S 4.2.5 how this may affect the optical-UV SEDs and resultant "ages" of tail sources derived by \citet{fuma11}.
The velocities of the HII regions and the redshifts of the background galaxies are indicated in
 Figure~\ref{fig11}.

\begin{figure}[htbp]
\epsscale{1.3}
\plotone{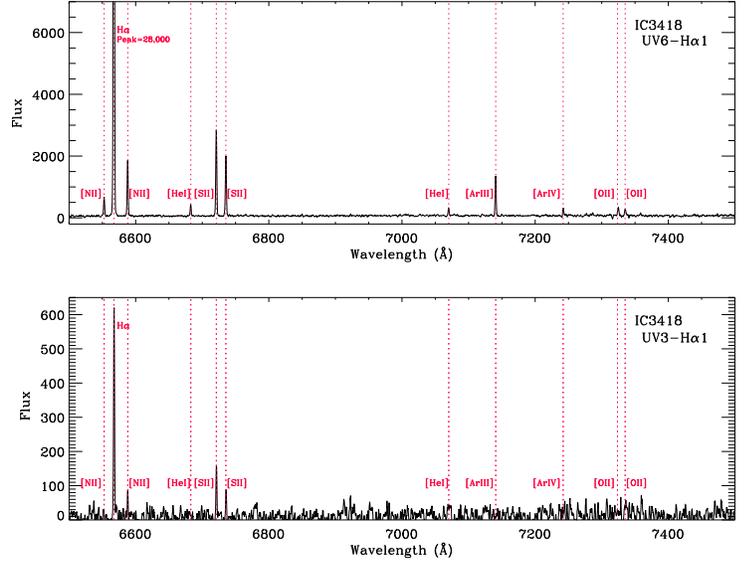} 
\caption{
Keck DEIMOS spectra of two of the HII regions in the tail of IC3418, with lines identified. The upper panel shows the strongest HII region UV6-H$\alpha$1, and the lower panel shows one of the faintest HII regions, UV3-H$\alpha$1, from the inner tail. The flux scale is in arbitrary units.
\label{fig13}
}
\end{figure}

Figure~\ref{fig13}
shows spectra for two of the HII regions. Multiple lines are detected for all HII regions, with line ratios characteristic of HII regions with sub-solar metallicity. The [NII]$\lambda$6583/H$\alpha$ ratios range from 0.06 to 0.12, which correspond to 12+log(O/H) ratios of 8.22 to 8.38, or 0.36-0.53Z$\solar$ (within a factor of 2.5), according to \citet{pett04}.  Thus the gas in tail has a metallicity similar to the stars in the main body.

\begin{figure}[htbp]
\includegraphics[scale=.38,angle=90]{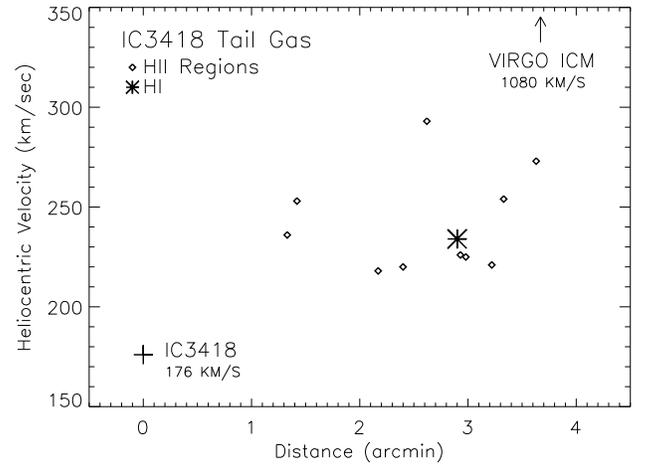}
\caption{
Position-velocity diagram for gas in tail of IC3418. Velocities of HII regions and the HI cloud plotted versus distance from the center of IC3418.
\label{fig14}
}
\end{figure}

The velocities of tail HII regions are shown
in a position-velocity diagram in  Figure~\ref{fig14}. The velocities of all tail HII regions (218-293 km/s) are redshifted with respect to the main body of galaxy (V=176 km/s). This is consistent with acceleration by ram pressure, since the the tail sources have velocities which are offset toward the cluster velocity.  
An inspection of  Figure~\ref{fig14} shows that a velocity gradient can be fit from the central velocity of IC3418 through the tail HII regions, but random motions are comparable in amplitude to any gradient. A large range in the velocities of young tail stars is also seen in simulations of ram pressure stripped tails \citep{tb12}, presumably reflecting the fact that clouds with different initial gas surface densities will experience different accelerations. The simulations also show a clear gradient in velocities, but the tail in IC3418 may not be long enough (and extend to high enough velocities) to show such a clear gradient. The tail kinematics should also reflect the disk kinematics. A lateral gradient across the tail is expected from the disk rotational motion, and has been observed in the tail star clusters of the stripped spiral galaxy ESO 137-001 \citep{sun10}.  There is no clear lateral gradient among the tail HII regions in IC3418, 
probably because the rotation curve is shallow in dwarf galaxies, and the velocity variations are dominated by other factors.


The tail HII regions have velocities which are low relative to the galaxy, and much closer to the galaxy (v=176 km/s) than the cluster ICM (V=1080 km/s).  This helps explain the fireballs, i.e., HII regions at heads of linear UV-bright stellar streams.  Since the velocities of the tail HII regions are still very different from that of the cluster ICM, they are probably still experiencing strong ram pressure from the cluster ICM,  and strong gas acceleration from ram pressure is required to cause the observed spatial offsets between the gas and stars in the fireballs.  

The simulations of \citet{tb12}  predict stripped gas tails with a
large range of velocities at any distance, due to differential stripping and acceleration. 
These simulations show that some gas extends to the cluster velocity, but even after a few hundred Myr of stripping, most of the gas is at velocities less than the cluster velocity, and much of the gas has velocities 1/3-1/2 of the way to the cluster velocity. 
However the tail HII regions in IC3418 extend kinematically much less than this, only 15\% of the way to the cluster velocity. Thus either the gas tail is much longer and young stars exist only in the inner part, or the tail is dynamically young and no part of it has had sufficient time to be accelerated closer to the cluster velocity.
The former is more likely in IC3148, given the absence of gas in the main body which indicates a relatively advanced evolutionary stripping state.


The low velocities also imply that while some of the tail sources will likely escape and add to the intracluster stellar population, others will likely fall back into the galaxy and add to the galaxy's halo.  Assuming the mass model of IC3418 described by  \citep{jac13}, which has a total mass of 7.5$\times$10${^9}$ M$\solar$, the escape speed is $\sim$110 km/sec at r=5 kpc, the projected distance of the inner tail, and is  $\sim$60 km/sec at r=17 kpc, the projected distance of the outermost tail.  Due to projection effects, the tail sources have higher galactocentric distances than the projected ones and therefore lower escape speeds, and also have higher galactocentric (3D) velocities. HII regions in the inner tail have l-o-s velocities which differ from the main galaxy by 40-75 km/sec, and the 3D velocities are probably $\sqrt{2}$ higher, or 55-105 km/sec. Some of these velocities are less than the local escape speed of $\sim$100 km/sec, meaning that some of the inner tail features are likely to fall back into the galaxy.  However some of the HII regions in the outer tail have l-o-s velocities which differ from the main galaxy by 95-115 km/sec, and the 3D velocities are probably $\sqrt{2}$ higher, or 135-160 km/sec. Even their projected velocities exceed the local escape speed of $\sim$60 km/sec, meaning these sources are definitely unbound and will escape into intracluster space.

A recent paper by \citet{ohyama13} presents spectroscopy of an emission line star in the direction of the tail, and suggests that it is a supergiant located within the tail. The velocity of this star is -99 km/s, which is inconsistent with the velocities of the HI and HII regions detected in the tail,
which range from 218-293 km/s. It remains unclear whether this star is physically associated with IC3418 or is instead a foreground Milky Way star.

\subsubsection{Acceleration Timescales for Tail Sources}

We can relate the tail length and velocities to the timescale for accelerating the gas parcels which are forming stars in the tail. 
The line-of-sight and sky components of velocity and distance of a tail source with respect to the main galaxy, are related by

\begin{equation}
\begin{array}{l}
\displaystyle {\rm L_{\rm tail}(los) = f \, v_{\rm tail}(los) \, t } \\
\displaystyle {\rm L_{\rm tail}(sky) = f \, v_{\rm tail}(sky) \, t } \\
\end{array} 
\label{eq:xdef}
\end{equation}

\noindent 
where
L$_{\rm tail}$(sky) is the distance travelled in the plane of the sky,
L$_{\rm tail}$(los) is the distance travelled along the line-of-sight,
v$_{\rm tail}$(sky) is the current velocity in the plane of the sky, 
v$_{\rm tail}$(los) is the current velocity along the line-of-sight,
t is the time of acceleration, and 
f is a fraction that depends on the acceleration profile, and is a measure of the time-averaged acceleration.

If a cloud starts with zero velocity and the acceleration is constant, f = 1/2.
There are three reasons why the acceleration is not constant.
First, the acceleration is due to the fight between ram pressure and gravity, and 
the gravitational restoring force depends on the initial gas position in the galaxy, and 
decreases as the gas moves away from the galaxy.
Second, the ram pressure  $\rho$(v$_{\rm gas}$-v$_{\rm ICM})^2$ on stripped clouds drops as they become accelerated away from the galaxy velocity toward the ICM velocity v$_{\rm ICM}$. This is a small effect for the tail sources in IC3418 since they have velocities much closer to their original (galaxy) velocities than the ICM velocity.
Moreover, the first and second effects act in opposite directions, so partly offset each other.
Third, the ram pressure on the galaxy varies strongly with time, as the galaxy orbits through the cluster. Simulations of orbiting galaxies by \citet{jac13}
suggests it can change by a factor of 2 over 200 Myr timescales.
We do not know whether the ram pressure is currently increasing or decreasing, 
so we adopt f=1/2 corresponding to constant acceleration and recognize the uncertainty is a factor of $\sim$2.

If we adopt f=1/2 and 300 Myr for the acceleration time of the outermost tail sources, from the star formation quenching time in the main body, then 
v$_{\rm tail}$(sky) = 2 L$_{\rm tail}$(sky)/t = 2 (17 kpc) / 300 Myr = 113 km/s and 
v$_{\rm tail}$(sky) = 1.13 v$_{\rm tail}$(los). 
The tail length along the line-of-sight would be 
L$_{\rm tail}$(los) =  f v$_{\rm tail}$(los) t = 1/2 (100 km/s) (300 Myr) = 15 kpc, 
and the tail 3D length is 23 kpc. 
We can also use this to estimate the sky velocity of IC3418, 
v$_{\rm gal}$(sky) = 1.13 v$_{\rm gal}$(los) = 1.13(900 km/s) = 1000 km/s, since for small acceleration v$_{\rm tail}$(sky) / v$_{\rm tail}$(los) $\simeq$ v$_{\rm gal}$(sky)/ v$_{\rm gal}$(los). Thus it is likely that the sky and l-o-s components of motion for IC3418 are similar.\footnote{
It may be possible in the future to estimate one of the unknowns L$_{\rm tail}$(los) or v$_{\rm tail}$(sky), if the other becomes known, without knowing either f and t, by eliminating t from Equation 1. We have 
L$_{\rm tail}$(sky)*v$_{\rm tail}$(los) = L$_{\rm tail}$(los)*v$_{\rm tail}$(sky) = 1700 kpc km/s, since we know  L$_{\rm tail}$(sky) = 17 kpc and v$_{\rm tail}$(los) = 100 km/s for the outermost tail HII regions. While we do not know L$_{\rm tail}$(los) or v$_{\rm tail}$(sky) individually, we know the product L$_{\rm tail}$(los) * v$_{\rm tail}$(sky) = 1700 kpc km/s.
}

On the other hand, we can use these relationships to provide an independent measure of the quenching time, by requiring that the velocity of IC3418 is within a reasonable range. In order for IC3418 to have a 3D velocity less than 2000 km/s, which is the approximate maximum of any Virgo galaxy, 
v$_{\rm gal}$(sky)<1800 km/s, which corresponds to t>200 Myr. This agrees well with the quenching time measured from the stellar population in the main body, and serves as an independent measure of it. If the acceleration time were as short as 100 Myr, this would imply v$_{\rm gal}$(sky)=3000 km/s, which is unreasonably high.

\subsubsection{Ages of Stellar Features in Tail}

The stellar features in the tail exhibit a range of UV-optical colors.
There appears to be an overall color gradient in the tail, with 
the outermost tail sources being the bluest and youngest \citep{fuma11}.
There is also significant scatter about this mean radial trend in color.
Part of this is likely due to the inclusion of contaminating background galaxies (and possibly foreground stars),
and part of it may be intrinsic.

\citet{fuma11} estimated the ages of the tail stellar features from UV-optical SEDs, and found a large age range, from 80-1400 Myr. 
Such a large range would suggest a surprisingly long period of strong ram pressure stripping, and even calls into question a
ram pressure stripping origin for the stellar features in the tail, since 
some of the proposed ages are much older than the quenching time in the disk of 200-400 Myr.
Such old stars in the tail would require that some gas was stripped from the galaxy long before most of the gas was stripped -- either the outer disk gas was stripped long ($\sim$1 Gyr) before the inner disk,
or lower density gas from the inner disk was selectively stripped, decoupling from denser gas
which was not stripped for another $\sim$1 Gyr. The long timescales for both scenarios are problematic.
Orbital times in the cluster are $\sim$2-3 Gyr, during which ram pressure is expected to vary by 1-2 orders of magnitude,
depending on the orbit. 
Simulations show that most stripping occurs when the ram pressure increases significantly during the period of radial infall, an interval which generally lasts less than a few hundred Myr \citep{vollmer01}, much less than the 1.4 Gyr required from the stellar age estimates.

Alternatively, it may be that some of the ages are overestimates.
We note that the \citet{fuma11} photometry was measured in 6$''$ apertures, to match the GALEX resolution.
At this resolution, background and foreground sources could be important, especially in the optical. Whereas in the UV
the tail sources are much brighter than anything in the background, in the optical the tail sources are hard to pick out among the many background sources (Figure~\ref{fig1}). Spectroscopy of tail sources from Keck shows that this is indeed a problem.  Background galaxies, with redshifts between z=0.23-0.68, are confirmed to exist in some of the tail filaments. Their locations are shown in Figure~\ref{fig11} and given in Table 3.
\citet{fuma11} estimated ages of 600-1400 Myr for 5 of their 12 tail sources.
Of these, three (F3, F4, K4) have confirmed background galaxies, and in two of them (F3 and F4), the galaxy is the brightest optical source. Inclusion of these bright optical background sources in a UV-opt SED will clearly result in an overestimate of the stellar age.
Other tail features may also include background galaxies, but we did not position slits on many of the knots.
Seven of the 12 tail sources studied by \citet{fuma11} have age estimates of 80-390 Myr, which seems reasonable given the quenching time in the galaxy, and we suspect that all the tail sources have ages within this range.
Sorting out the true age distribution in the tail will require a more complete identification of true and contaminating tail sources.

While some of the color variation in tail sources is due to contaminating sources,  some intrinsic local variation in age may be expected in the tail. In AMR simulations, \citet{tb12} find that
disk gas  with a range of densities (although not dense clouds) is stripped continuously, so there is a range of gas densities throughout the tail. Accordingly, the time it takes for radiative cooling and possibly compression by the ICM, and consequent star formation to occur, also varies throughout the tail. Thus at any one time the tail may have dense gas and star formation over a large radial range.
Moreover, with an extended period of gas stripping, 
in the inner tail it is possible to have outwardly moving young stars at the same distance as older backfalling stars.
In the case of IC3418, which is now (almost) completely gas-stripped, there is no more gas from the galaxy to supply to the tail, and the gas in the tail has moved further downstream.
This may explain why the only bright HII regions and the bluest stellar features are in the outer half of the tail.


\subsection{Gas Content of the Main Body and Tail}

The main body of the galaxy is very gas poor 
as HI, H$\alpha$, and X-Ray emission are undetected
\citep{hof89,chung09,jac13}.
There is a marginal detection of CO in the central few arcseconds of galaxy, 
corresponding to a gas mass of $\sim$10$^6$ M$\solar$, 
with the standard CO-H$_2$ relation \citep{jac13}.
Even if this is a real detection, the amount of gas is very small.
IC3418 is HI-deficient and H$_2$-deficient compared to star-forming dwarfs \citep{jac13},
with less than 1$\%$ of the expected HI+H$_2$ mass.
Virtually all the gas has been stripped from its main body.

The only gas detected in IC3418 is in the tail.
In addition to the HII regions, there is 3.7$\times$10$^7$ M$\solar$ of HI.
CO is undetected in the tail with an upper limit which corresponds to an
H$_2$ mass <10$^7$ M$\solar$, assuming a standard CO-H$_2$ relationship \citep{jac13}.
The gas consumption timescale due to star formation (also known as the inverse of the "star formation efficiency"),
computed using star formation rate of 1.9$\times$10$^{-3}$ M$\solar$ yr$^{-1}$ from H$\alpha$,
and the sum of HI+H$_2$ masses, is 2$\times$10$^{10}$ yr .
This timescale is similar to that in the stripped extraplanar gas of 3 Virgo spirals  \citep{vollmer12},
and several times longer than the timescale in typical spiral disks.
This long timescale (or low "efficiency") indicates that most of the stripped gas does not form stars, but remains gaseous and ultimately joins the ICM.

X-Ray and diffuse H$\alpha$ emission are undetected in the tail \citep{jac13}.
The upper limits on mass depend on the filling factors assumed,
but for reasonable choices, the upper limits are 
4$\times$10$^7$ M$\solar$ for X-Ray emitting hot gas 
and 6$\times$10$^7$ M$\solar$ for H$\alpha$-emitting warm diffuse gas \citep{jac13}.
The lack of X-Ray emission might be partly due to a
low thermal pressure in Virgo, compared to other clusters where
X-Ray tails have been observed, resulting in a lower X-Ray emissivity for the gas \citep{tbc11}.
On the other hand, the tail does seem to have a low gas content.
It could be in an advanced evolutionary state, as
there is no longer any gas from the galaxy to supply the tail.
There may be gas located downstream from the observed young stars in the tail,
a real possibility given the low velocities of HII regions in the tail.

\subsection{Orbit within the cluster}

\begin{figure}[htbp]
\epsscale{1.1}
\plotone{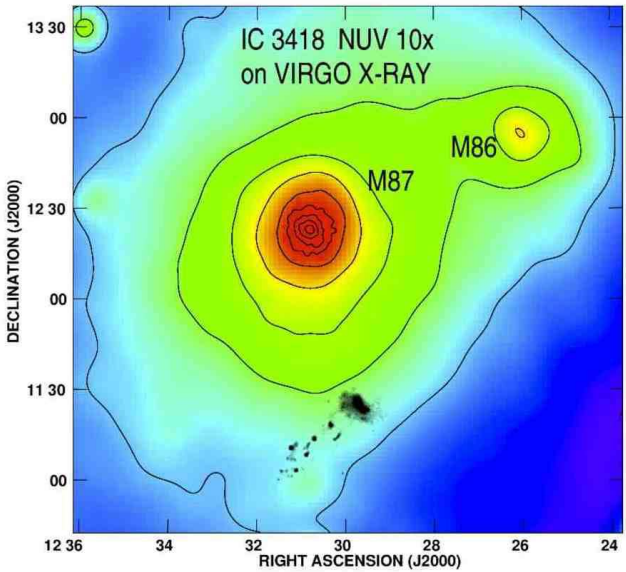}
\caption{
GALEX NUV greyscale image of IC3418, increased in scale by factor of 10,
on ROSAT X-Ray contour map of Virgo cluster.
Contour increment is factor of 2.
The position of the nucleus of the main body of IC3418 is correct.
\label{fig15}
}
\end{figure}

The small projected radial distance of IC3418 from M87 and the presence of a gas-stripped tail support the idea that IC3418 may be near its closest approach to M87. IC3418 is moving toward us with respect to the cluster center at ~900 km/sec.
This moderately high line-of-sight velocity implies that the galaxy 
is located somewhere between the cluster center and intermediate clustercentric distances, but not near apocenter, where 
a near-zero line-of-sight velocity would be expected.

Figure~\ref{fig15} shows a NUV image of IC3418 on a map of the Virgo cluster.
If ICM motions are small, the tail indicates the projected direction of motion of IC3418 through the cluster. Figure~\ref{fig15} shows that the tail position angle of 132$\deg$ is offset by 115$\deg$ from the vector connecting IC3418 and M87. The component of orbital motion in the plane of the sky is thus $\sim$2 times more tangential than radial.
This is different from
several Virgo spiral galaxies at larger projected distances (500-800 kpc), which have gas tails pointing roughly radially away from M87 (Chung et al. 2007). These are thought to be galaxies on highly radial orbits which are falling into the cluster for the first time. Since their projected tails have large radial components, they are likely far from closest approach. 
In contrast, the projected tail angle in IC3418 is more tangential than radial.
If this is also true for the 3D tail angle, then IC3418 would be near closest approach,
although projection effects can be misleading.

We have calculated cluster orbits consistent with the observed orbital components of IC3418: the line-of-sight velocity,
and the plane of sky position and direction of motion, the latter inferred from the tail angle.
The free parameters of the models are the current line-of-sight position of IC3418 and the amplitude of its plane of sky velocity. 
We assume a smooth and spherically symmetric mass distribution for the Virgo cluster, represented with a radial $\beta$-profile
(Schindler etal 1999).  The cluster mass distribution is truncated at 2 Mpc distance. 
Based on the likelihood of IC3418's location near pericenter, as explained above, we have varied the current line-of-sight position relative to M87 in the interval of (-500, 500)~kpc. The sky plane velocity varied from 700 to 1400~km/s, resulting in radial orbits with peri-to-apocenter ratios changing from about 1:5 to 1:20. Velocities outside this range would yield either too compact or too prolonged (close-to unbound) orbits. Fast orbits at large distances from M87, or slow orbits close to M87 are also less probable from the point of view of distribution of orbits in galaxy clusters (Benson 2005). 
All consistent orbits have IC3418 within about 350~Myr of closest approach to M87.
While the observed orbital components and the simulated orbits cannot determine whether it is currently before or after pericenter passage, it is statistically more likely to have a line-of-sight position relative to M87 of $\leq$300 kpc than a distance as large as
500 kpc.

Additional orbital constraints come from the observation of recent quenching in IC3418.
Since most of the gas removal from the disk occurs before pericenter passage, the quenching time of 300$\pm$100 Myr in the center of IC3418 suggests that the galaxy is presently no more than 400 Myr after pericenter, and possibly still before pericenter.
The presence of HI at the observed position in the tail, only 14 kpc from the main body, suggests that IC3418 is in a pre-pericenter stage. The modelling of \citep{jac13} suggests that parcels of gas with column densities typical for HI would be stripped to much larger distances after pericenter.
The presence of fireballs, which are unusual in the Virgo cluster, suggests an extreme ram pressure for Virgo, and this also argues for a current position close to the core.

An upper limit on the current ram pressure felt by IC3418 is $\sim$1800~cm$^{-3}$(km/s)$^2$, 
assuming a smooth static ICM, a sky plane velocity of 1400~km/s, 
and a distance from M87 equal to its projected distance. If the adopted position angle PA=50$\deg$  and inclination i=30$\deg$ of IC3418 are correct, the wind angle is currently within 25$\deg$ of face-on, for all consistent orbits, if the SE side of the disk is the near side.
The 3D length of the tail is 1.2-1.7 times larger than its projected length, if the plane of sky velocity is between 
the reasonable limits of 700 to 1400~km/s, as suggested above. 

Ram pressure of this strength is sufficient to strip gas with a surface density of 15 M$\solar$ pc$^{-2}$ from the center of IC3418, 
or 50 M$\solar$ pc$^{-2}$ at a distance of 5 kpc from the center, 
but is insufficient to strip GMCs with surface densities of $\sim$100 M$\solar$ pc$^{-2}$ \citep{jac13}.
Thus dense clouds in the disk of IC3418 likely decoupled from surrounding lower density gas which was stripped, and then were either ablated by Kelvin-Helmholtz instabilities, or evolved to a lower density state that was directly stripped \citep{crowl05, ak13}.
The dense star-forming clouds in the tail would not have originated from disk GMCs which were directly stripped but instead formed locally in the tail via cooling \citep{tb12}.
Because of the much lower gravitational forces in the tail, GMCs which form in the tail can be directly accelerated by ram pressure, and this likely plays an important role in the fireball phenomenon.


\section{Discussion}


\subsection{Fireballs and Linear Stellar Streams}

The positions of the HII regions at the outer heads of 
UV-bright head-tail and linear stellar streams ("fireballs")
strongly suggest a scenario in which
star-forming gas clouds continue to be accelerated outwards by ram
pressure, depositing newly-formed stars behind it on its path. 
The stars are unaffected by the ram pressure, 
so the stars separate from the gas cloud.
Once stars decouple from gas, they may fall back into the galaxy following ballistic orbits \citep{hester10}
if their birth cloud had not achieved escape velocity.
Such fallback stars would add to the galaxy's halo or thick disk \citep{abram11},
but not principally to its bulge, as was suggested by \citet{kap09}.

\begin{figure}[htbp]
\plotone{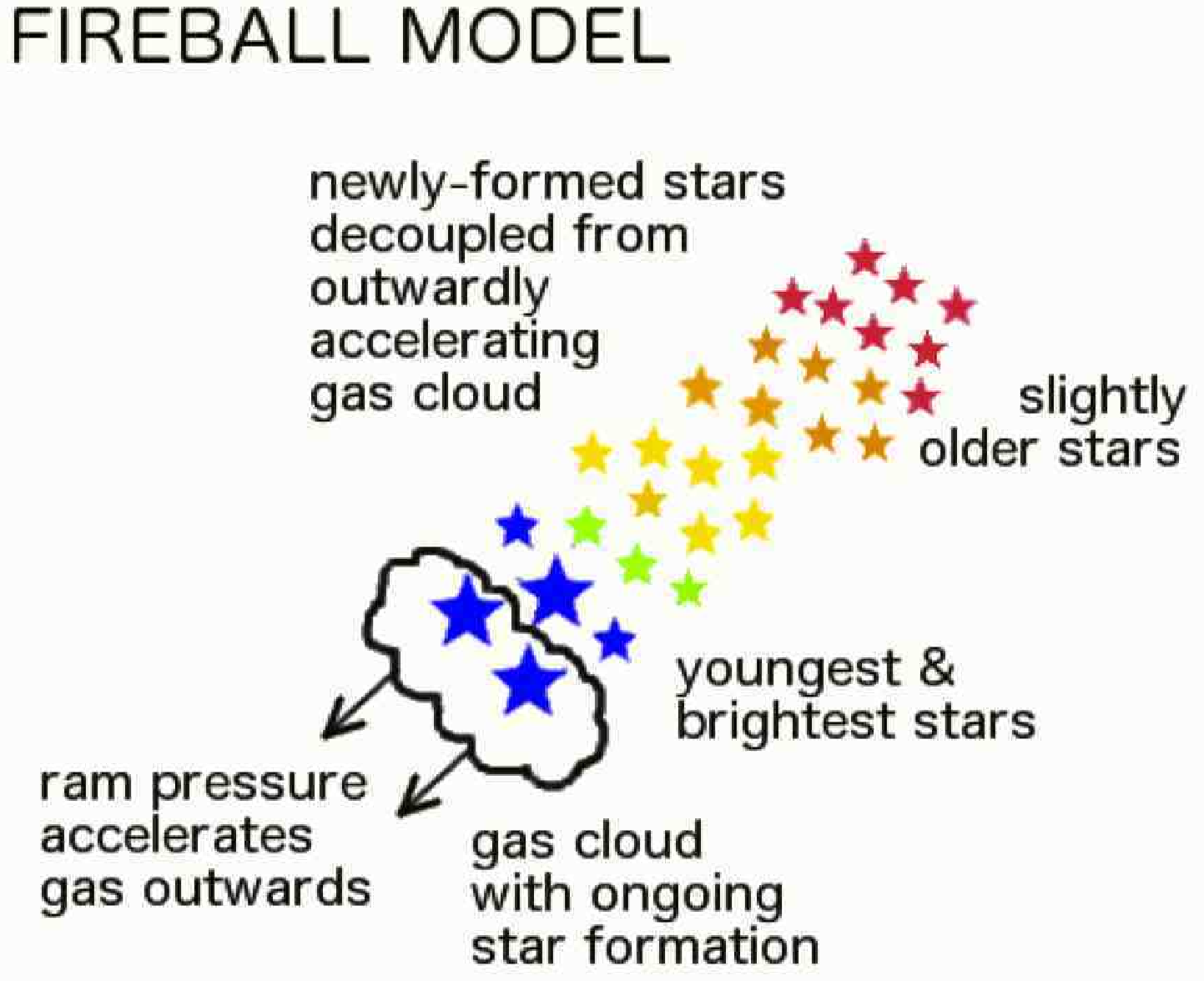}
\caption{
A cartoon of our fireball model.
A gas cloud is accelerated outwards by ongoing strong ram pressure.
It is trailed by young stars which formed in the gas cloud, but are now decoupled from the gas since the stars don't feel the ram pressure.
The stellar associations nearest the gas cloud are the youngest, so still contain the most massive and luminous stars, and this forms the luminous head of the head-tail source (or "knot"). 
Further from the gas cloud are slightly older stellar associations which no longer contain the most massive and luminous stars, and this forms the tail (or "filament"). The color differences of the stars along the tail are greatly exaggerated.
\label{fig16}
}
\end{figure}

We show a cartoon of our fireball model in Figure~\ref{fig16}.
It shows an outwardly accelerating gas cloud, trailed by decoupled young stars.
The stellar associations nearest the gas cloud are the youngest, so still contain the most massive and luminous stars,
and this forms the luminous head of the head-tail source (or "knot"). 
Further from the gas cloud are slightly older stellar associations which no longer contain the most massive and luminous stars, and this forms the tail (or "filament"). 
Followup studies are needed to confirm this predicted stellar age gradient in the head-tail sources.  
We note that within a galaxy, the large gravitational restoring force can prevent dense star-forming GMCs from being directly stripped by ram pressure. But far from the galaxy in the tail, the gravitational restoring force is weak,
and this makes it easier to accelerate star-forming GMCs by ram pressure.


Stars that form in the tail will have the same initial velocity as the parental gas cloud, and both are subject to the same gravitational acceleration from the galaxy.  The only difference is that the gas clouds continue to be accelerated by ram pressure, but the stars are not.
Thus the timescale t$_{\rm stream}$ for creating a linear stellar stream of length L by ram pressure acceleration a$_{\rm ram}$ of a (decoupled) gas cloud of surface density $\Sigma _{\rm gas}$ is 
${\rm t_{\rm stream} = (2L/a_{\rm ram})^{1/2} }$ \citep{hester10}, or 

\begin{equation}
\displaystyle {\rm t_{\rm stream} =  {\sqrt{\frac{2L \, \Sigma_{\rm gas}}{\rho_{\rm ICM} \, v^2}}}}
\label{eq:tstream}
\end{equation}

\noindent
For a GMC-like gas surface density of $\Sigma _{\rm gas}$ = 100 M$\solar$ pc$^{-2}$, 
an ICM density $\rho _{\rm ICM}$ of 10$^{-3}$ cm$^{-3}$, and a relative velocity v of 1400 km/s,
it takes t$_{\rm stream}$ = 60 Myr to create a stellar stream with L = 1 kpc.
With such an age difference between the head and the tail, 
a gradient in optical/UV colors should be detectable with sufficient sensitivity.
For the linear stream UV3, which has a length of $\sim$2 kpc, 
\citet{hester10} measure a UV color gradient 
corresponding to an age gradient of $\sim$100 Myr. This is in good agreement with our estimate.
Stellar streams are expected to be longer in clusters with higher ram pressure.
This seems to be the case for the Coma galaxy RB 199, which has some streams as long as 10-20 kpc,
\citep{yos08}, much longer than those in IC3418, presumably because the ram pressures in Coma are much higher than in Virgo \citep{hester10}.
Streams are generally longer in the inner tail of IC3418 than the outer tail.
Two effects may contribute to this difference: the outer tail streams are still forming, and the inner tail streams are tidally stretched, since close to the disk the gradient of restoring force along the filament is larger than at larger z-distances.

An alternative way to make linear stellar streams is to make linear streams of dense gas by ram pressure.
This could happen via differential acceleration of gas elements with different densities, forming a linear gas stream which then cools and forms stars along its length.
This process may happen in some galaxies. 
RB199 exhibits both head-tail features like those in IC3418,
with relatively compact H$\alpha$ concentrations at the head of a linear stream of young stars pointing back toward the galaxy,
but also some linear H$\alpha$ features as long as 18 kpc.
The linear H$\alpha$ features in RB199 are preferentially in the outer part of the tail, and are fainter in the UV-optical than the inner tail features, suggesting that they are lower mass features which form a small fraction of the tail stars.
We attribute the differences between RB199 and IC3418 to stronger ram pressures in the Coma cluster.

\subsection{How IC3418 compares to other galaxies with ram pressure stripped tails}


Several Virgo spiral galaxies have tails of HI gas produced by ram pressure
stripping
\citep{kvgv04,oovg05,crowl05,chung07,abram11},
and some of these have
UV and H$\alpha$ sources from recent star formation in the tails
\citep{kk99,cor04,crowl05,abram11}.
But unlike IC3418 none of these Virgo spirals are completely
gas-stripped, none have such a luminous UV tail, and none have fireballs or linear stellar
streams produced by ram pressure.
The Virgo spiral NGC~4330 is similar to IC3418 in having 
an extraplanar UV tail, and HII regions only in the outer part of the UV tail \citep{abram11}.
But IC3418 differs from NGC~4330 in having a symmetric UV tail, as well as fireballs in the tail.
IC3418 is $\sim$1-2 mag fainter and has 10 times less stellar mass than the 
prototypical stripping-in-action Virgo spirals NGC 4522, NGC 4402 and NGC 4330
\citep{kvgv04,crowl05,abram11}. 
The spirals in Virgo may be too massive to be so completely and dramatically
stripped. The dwarf IC3418 is both lower in mass and closer to the cluster
center (at least in projection)\footnote{
Located 1.0$\deg$ from the cluster center,
IC3418 is closer to the cluster center, in projection,
than all the Virgo spirals known with ram pressure stripped tails,
which are located at 1.3-3.5$\deg$ from the cluster center. 
It has the same projected distance as the gravitationally disturbed spiral NGC~4438,
which may be experiencing ram pressure stripping from the cluster \citep{vollmer09}
but has also experienced a damaging high-velocity collision with M86 \citep{ken08},
so its case is complicated.
}
than the other Virgo galaxies known to be experiencing ram pressure stripping.

The phenomenon of one-sided tails of young stellar knots and parallel
linear stellar streams has been seen in more distant clusters.
Numerous galaxies with one-sided tails of H$\alpha$ emission or young stellar knots and parallel
linear stellar streams have been observed in the Coma cluster \citep{yos08,yagi10,smith10,fos12}.
\citet{cor07} have described two peculiar galaxies falling into the massive
z$\sim$0.2 galaxy clusters Abell 1689 and Abell 2667, each with 
extraordinary 1-sided tails of bright blue knots and stellar streams.
Two spiral galaxies in the Norma cluster (ESO137) have
very prominent X-Ray and H$\alpha$ ram pressure stripped tails \citep{sun10}.
We think all these galaxies are observed in a short-lived phase of very strong
ram pressure stripping in which a large fraction of the ISM is stripped out.
These galaxies are all more massive than IC3418, but
their clusters are also more massive, with denser ICMs than Virgo, and have
peak ram and ICM thermal pressures $\sim$10-100 times that in Virgo.
It may be that rapid strong stripping can occur for large galaxies in massive
clusters, but only for dwarf galaxies in less massive clusters like Virgo.


Features which resemble linear stellar streams are seen in the simulations of
both \citet{kap09} and \citet{tb12}, and
head-tail features, with young stars at the heads of elongated stellar streams,
are clearly seen in the \citet{kap09} simulations (although not commented on in their paper).
Both the ram pressure and thermal pressure seem to be important for producing the 
star formation in the tails and the linear stellar streams.
\citet{kap09} and \citet{tb12} both find that the amount of star formation in the tail
is higher when the ambient thermal ICM pressure is higher,
presumably since gas clouds experiencing external pressure can collapse more readily.
Both increase toward the center of the cluster, and this could be why IC3418 but not
the Virgo spirals, at larger projected distances, show pronounced stellar streams.
The lower gravitational potential of IC3418 may also be relevant, since with the same ram
pressure its gas can be more easily accelerated and stripped.




\subsection{Other Ram Pressure Stripped Virgo dwarfs}

There are other Virgo dwarf galaxies which have been proposed to be ram pressure stripped.
The dwarf UGC7636=VCC1249, located very close to the large Virgo elliptical M49,
is likely experiencing the combined effects of tidal and ram pressure stripping 
due to a collision with M49 \citep{mcnam94,lee00,arr12}.
Optical images show the outer parts of the dwarf's stellar body are distorted in a way consistent with a tidal interaction, while the morphology of the inner region is bow-shaped at the apparent leading edge, suggestive of ram pressure.
There is no HI in the stellar body but there is an HI cloud nearby, between M49 and the dwarf in both space and velocity. The dwarf's gas has likely been stripped by ram pressure from the gaseous halo of M49.
While an excellent case for ram pressure stripping, UGC7636 is stripped by the halo of M49 rather than the general Virgo ICM, and since it passes close to M49, it has experienced a strong tidal interaction that accompanies the ram pressure stripping.
Most cluster dwarfs are likely stripped by the general ICM rather than the ISM of galaxies.

There are a few Virgo dIs, detected but deficient in HI, which seem partially stripped, and are
candidates for active stripping by the cluster ICM.
The Virgo dI IC3365, located 3.5$\deg$ from M87, shows evidence for ongoing ram pressure stripping. Its HI content is $\sim$5-10 times lower than most dIs, with M$_{\rm HI}$/L$_{\rm B}$ = 0.3, and its HI contours seem compressed on one side \citep{skill87}.
\citet{lee03} identified 5 Virgo dIs with lower gas mass fractions at a given oxygen abundance than non-cluster dI's. These galaxies are HI-deficient by factors of 7-30, and are among the dIs located closest to the cluster center.
While these are good candidates for active stripping, none are
known to have tails of only gas and young stars, which is a clear signature of active ram pressure stripping.

There are numerous early type dwarfs which are candidates for post-stripped galaxies.
In an early study \citet{vig86} proposed that the large dwarf IC3475
with an uncertain classification (ImIV or dE1 pec according to \citet{bst85}) 
is ram pressure stripped.
It has no HI or ongoing star formation (H$\alpha$) detected to low limits, 
although it has substructure in the disk (a bar and knots) and intermediate optical colors which are
bluer than most dEs (B-R=1.5), but 
sufficiently red that star formation must have stopped $\geq$1 Gyr ago.
It is located very close (in projection) to the cluster center, only 35$'$ from M87,
although if the quenching time is $\geq$1 Gyr, then it must have been stripped before current core crossing. 
It remains an excellent candidate for a previously stripped dwarf,
and may be similar to dEs with low-level disk substructure.
\citet{jer00} discovered spiral structure in the dE IC3328, which could be a gas-stripped dI.
Nearly 10\% of 476 Virgo dEs in an SDSS imaging study 
show low-level disk substructure including spiral arms, bars, and edge-on disks \citep{lis06a}, 
and roughly 5\% show blue centers indicating recent star formation \citep{lis06b},
with much higher fractions for the brightest dEs.
Both populations are consistent with being gas-stripped dIs.

While these galaxies are good examples of likely ram pressure stripping of dwarfs in Virgo,
IC3418 is an even better example because it has clearcut evidence for active ram pressure,
it is (almost) completely stripped, and it is stripped by the Virgo ICM.

\subsection{DIs->dE's by ram pressure stripping}

We propose that IC3418 is a dI which was very recently almost completely stripped of its gas by
ram pressure. As its last generation of stars fades and reddens, and substructure
within the stellar disk disperses, the galaxy will presumably become a dE.

Many previous studies have concluded that ram pressure stripping of infalling dIs is a likely origin, or partial origin, 
for cluster dEs \citep{lee03,vanzee04a,vanzee04b,bos08,kb12}.
dEs and dIs have similar masses and follow similar scaling relationships for their luminosity and
radial stellar light distributions,
including the shape of the radial profiles (close to exponential),
central surface brightnesses, and scale lengths \citep{lf83, lee03,kb12}.
The parameter correlations of absolute magnitude, effective radius, and surface brightness 
are very similar for dIs and dEs, and quite distinct from ellipticals \citep{kb12}.
Dwarf galaxies (10$^7$ < M < 10$^9$ M$\solar$) with no active star formation are extremely rare in the field,
and the fraction of quenched dwarf galaxies decreases rapidly with increasing distance from a massive host
\citep{geha12}.
This indicates that some interaction in a dense environment,
rather than any internal mechanism, is responsible for quenching star formation in dwarfs.

IC3418 follows the scaling relation for other dIs and dEs.
Given its exponential light profile with scale length of 19$''$=1.5 kpc,
and its central surface brightness of 23 B mag arcsec$^{-2}$, and its B magnitude of
-16.5, it falls in with other dwarfs in Figure 19 of \citet{lee03}.


Thus the faded remnants of IC3418 and other gas-poor dI's in Virgo should closely resemble at least some of the dEs currently seen in the cluster core \citep{lee03}, particularly the dEs with blue centers \citep{lis06b} or disk substructure \citep{lis06a},
Blue centers are naturally produced by non-instantaneous ram pressure stripping, since
galaxies are stripped from the outside in.
While some of the blue center dEs could also have had tidally triggered central star formation,
their lack of gas is likely due to ram pressure stripping.
IC3418 has a color gradient of 0.15 in g-r \citep{hester10},
which exceeds that required to be defined a blue centered dE by \citet{lis06b},
so it could evolve to become one of blue center dEs.
Since it has spiral structure and a disk, it could also evolve to become
one of the disk substructure dEs of \citet{lis06a}.

Objections to the ram pressure stripping scenario for the origin of cluster dEs are based on
differences between the dI and dE populations that cannot be accounted for by ram pressure alone,
including kinematics, shapes, metallicity, and nuclei \citep{fb94,con01}.
These real differences indicate that tidal interactions and perhaps other mechanisms 
have also played an important role in the evolution of most dEs \citep{kb12}.

For example, ram pressure stripping cannot explain the observed differences in stellar kinematics
between these galaxy classes.  dIrrs are rotationally supported, with v/$\sigma$$\geq$1,
whereas dEs have heterogenous kinematics. Some have rotation and some don't, and
all known dE's have v/$\sigma$$\leq$1 \citep{geha03,vanzee04a}.  
These kinematic differences seemingly require gravitational interactions for at least
some of the dEs. 
Numerical simulation have suggest that 'galaxy harassment'
\citep{mlk98,mas05}, the accumulation of gravitational heating via multiple
minor encounters, can transform a rotationally supported galaxy into a dispersion
supported galaxy (although the importance of this effect has been questioned by \citep{sdn10}.)
Careful photometric analysis by \citet{kb12} shows that
some Virgo dEs have clearly puffed-up outer disks consistent with tidal interactions .

Many dEs have nuclei \citep{cote06}, which are rare in dIs of the same mass \citep{bts87,fb94}, although
common in more massive Sd and Sm galaxies \citep{boker04,kb12}.
Whereas dIs and  non-nucleated dEs have the same shape distributions, 
nucleated dEs may be rounder on average \citep{fs89,lis07}.
These facts suggest that the progenitors of nucleated dEs are more massive Sd and Sm galaxies, and that the dE has lost mass and become rounder over time due to tidal interactions \citep{kb12}.

Until recently it was thought that
dSphs and dEs had somewhat higher metallicities than dIs of the same luminosity or mass,
suggesting that dSphs had undergone comparatively 
more early and rapid star formation episodes \citep{mateo98,gre03,fw11}.
However this was based on photometric estimates of metallicities, which can suffer from systematic errors.
A newer study based on spectroscopic metallicities shows that dIs and dSphs in the Local Group obey the same mass-metallicity relation \citep{kir13}.
Similar mass-metallicity relations for dIs and dEs agrees with the results of  \citet{weisz11},
who find similar star formation histories for nearby early and late types dwarfs.
They find that the average dwarf formed $\sim$50\% of its stars by z$\simeq$2 and $\sim$60\% by z$\simeq$1, regardless of morphological type.
The mean star formation histories of dIs, transition galaxies, dSphs are similar over most of cosmic time, and only began to diverge a few Gyr ago, with the clearest differences appearing during the most recent 1 Gyr \citep{weisz11}. We note that identical mass-luminosity relations for all dIs and dEs might not be expected, if some dEs have lost mass through tidal interactions, as seems likely.



dEs are a heterogenous population with a range of evolutionary histories \citep{lis07}, and 
tidal interactions have clearly helped shape a large fraction of dEs.
But the one thing they have in common is a lack of gas and star formation,
and this is probably caused in most cases by ram pressure stripping, since
tidal interactions do not selectively remove gas.

It is clear that ram pressure stripping is not the only interaction which has affected most of the cluster dEs.
Simulations of local group dwarfs indicate that both ram pressure stripping and tidal interactions need to occur 
to account for all properties \citep{may07}.
In the dense cluster environment it is only natural, and even expected, that most galaxies should experience
both ram pressure stripping and tidal interactions, \it although generally at different times. \rm
For most cluster dwarfs, removal of all or most of the gas probably
occurs during one ram pressure stripping event during its first approach to the core.
Gas removal is generally a single event localized in time and space, 
and is decisive for the dI->dE transformation.
In contrast, tidal interactions can occur as pre-processing before the first infall, 
during the first infall, and on any subsequent orbit.
Tidal damage and mass loss can increase through multiple interactions,
be widely spread over time and cluster locations and orbits, and
significant tidal damage can occur after the first infall, when gas is stripped.
Although tidal interactions are important for dwarf galaxy evolution 
and responsible for some dE properties, they are not decisive for the dI->dE transformation.

dIs are those dwarf galaxies that have been in cluster the
least amount of time, and have never been through the cluster center.
These galaxies are expected to be much less tidally disturbed than dEs, on average.
Once a gas-rich dwarf enters the central part of a cluster like Virgo, 
it will be largely gas-stripped.
The widespread ram pressure stripping of spirals in Virgo, in which massive spirals 
that get within $\sim$0.5 Mpc of the cluster core, typically get stripped to 
0.3-0.7R$_{25}$ \citep{kk04b,chung09},
implies that dwarfs, with their shallower gravitational potentials,
should be completely or nearly completely stripped on its first core passage.

dEs are those dwarf galaxies that have been in cluster for longer than the dIs, 
and the subpopulation of dEs which are nucleated, centrally concentrated and slow-moving
may be the oldest subpopulation \citep{lis09}.
dEs in general have been fully (or nearly fully) stripped of their gas\footnote{
While most dEs are gas-poor and undetected in HI, a small fraction of dE's in Virgo (4-15\%) are detected in HI \citep{con03} \citep{hall12}, with values of M$_{\rm HI}$/M$_{\rm stars}$ similar to those of late type galaxies.
Nearly all have projected locations in the cluster outskirts. 
Of the 12 dEs detected in HI by the ALFALFA survey, 6 are blue and are likely misclassified dIs, 
6 are red and may be best explained by quiescent galaxies (possibly previously stripped) which have recently accreted HI
\citep{hall12}.
}
and have made one or more orbits in the cluster, subjecting themselves to multiple
tidal encounters with the cluster and other galaxies.
Such encounters tidally strip the outer galaxy, and dynamically heat the bound stars.
The gravitational disturbance histories of cluster dwarfs vary,
dominated by a few discrete events, and each event can be different, depending on
the separation and relative velocity at closest approach, and the angles and angular momenta of the interaction.
This may help explain why dEs are not a homogenous class \citep{lis06a,lis06b,lis07,lis08}. 
While there may be different pathways to make the different kinds of dEs, gas stripping is likely necessary for all.

While the lack of gas and star formation and associated substructure is not the only difference between dIs and dEs,
it is the single most important difference since it determines whether a galaxy is a dI or dE.
A galaxy which experiences gas stripping but not a tidal interaction would (ultimately) be classified as a dE or dSph,
whereas a galaxy which experiences a tidal interaction but not gas stripping would not be classified as dE or dSph.

We propose that IC3418 is caught red-handed in that critical part of the dI->dE transformation,
the removal of its gas by ram pressure stripping.

\section{Summary}

1. IC3418's tail is one-sided, straight, and centered on the main body of galaxy, unlike tidal tails.
There is no evidence of a smoother, older stellar component to the tail, as would exist in a tidal tail.
All detected stars are clustered and young, consistent with all the tail stars originating from star formation
within a ram-pressure stripped gas tail.
A deep optical image indicates no strong gravitational disturbance to the main stellar body,
further evidence that the tail is the product of ram pressure stripping of gas.

2. Many UV tail features are linear, parallel streams of young stars.
Some of them are head-tail features, with a knot of UV-bright stars at the outermost head of a linear stellar stream.
H$\alpha$ emission is located at the heads of these, outwardly offset by $\sim$80-150 pc from the UV peaks.
The head-tail (``fireballs'') and linear stellar features in the stripped tail are likely formed from dense gas clumps
which continue to accelerate through ram pressure, leaving behind streams of newly formed stars which are not affected by
ram pressure.

3. HII regions in the tail have line-of-sight velocities 40-115 km/s higher than the galaxy. 
This velocity offset is in the direction expected for ram pressure stripping,
as IC3418 is moving toward us through the cluster at 900 km/s.
The velocities are relatively low, much closer to the galaxy than the cluster ICM.
This implies that gas in the tail is still experiencing strong
ram pressure, which is likely necessary to form the head-tail features ("fireballs").
In simulations ram pressure stripped tails extend closer to the ICM velocity, suggesting
that the tail in IC3418 is much longer than the currently known tail, and that star formation happens only in the inner part of the tail.

4. Most of the outer tail features have velocities which exceed the escape velocity, and will join intracluster space.
Some of the inner tail features likely remain bound to the galaxy, and should ultimately fall back in.
Some of the infalling stars will be in the form of extended halo streams, which differ in character from tidal halo streams in
being clumpy, young, and with possible age gradients. One likely infalling halo stream is identified in the UV image. 
Compared to the other tail streams, it is closer to the galaxy but further from the tail centroid, slightly curved, 
redder in color, and of lower surface brightness.

5. We identify 3 distinct zones in the tail.
The inner tail has UV-bright linear streams and knots, but no head-tail features and very little H$\alpha$ emission.
The outer tail has UV-bright head-tail features, i.e., knots with H$\alpha$ at the heads of linear stellar streams ("fireballs").
The very outer tail has lower surface brightness diffuse UV features with associated H$\alpha$ emission, and the 
highest H$\alpha$/NUV ratio.
We propose the following interpretation for the different tail zones.
Nearly all the gas has been stripped from the inner tail, and some stellar features here may be falling back into galaxy.
The outer tail still has gas and ongoing star formation, with some gas large gas concentrations being outwardly accelerated, forming fireballs.
The very outer tail still has gas and ongoing star formation, but smaller gas concentrations, so less bright stellar knots.

6. While no HI has been detected from the main body of the galaxy,
a small amount of HI, 4$\times$10$^7$ M$\solar$, is detected from the outer tail.
This corresponds to only 6$\%$ of the expected HI mass of a late type dwarf with the mass of IC3418.
The gas consumption timescale due to star formation, computed from the ratio of HI mass to the H$\alpha$-based star formation rate, is longer than the typical values in galaxy disks,
indicating that most of stripped gas does not form stars but ultimately joins the ICM.
The mass of stars in the tail is only $\sim$1$\%$ of the stellar mass of galaxy.

7. Spectroscopy shows that some sources in the tail region are background galaxies. Light from these contaminating sources were probably included in 
the \citet{fuma11} SED analysis which concluded that the ages of stellar features in the tail ranged from 80-1400 Myr. We confirm that some of the features with the oldest derived ages have bright contaminating background galaxies, so that some of the ages derived by \citet{fuma11} are likely overestimates. The true age range is probably closer to that of the features with the youngest derived ages, 80-390 Myr. Such an age range is consistent with the star formation quenching time in the main body of the galaxy.

8. We have analyzed the stellar population of the main body of IC3418 as a function of radius through optical spectroscopy and UV photometry.
Assuming a star formation history with a constant star formation rate until the time of quenching, with a possible burst at quenching time,
we find the quenching time in the galaxy center was 300$\pm$100 Myr ago.
A starburst occurred near the time of quenching, with a global average of 10$\pm$5$\%$ of the stars formed in the burst, although this varies with position in the galaxy.
Older quenching times at larger radii are expected from ram pressure stripping, as galaxies are stripped from the outside in.
The FUV/NUV ratio in IC3418 suggests a modest radial gradient in quenching time of 70 Myr from 0-48$''$,
indicating that it was stripped rapidly.
Stars in the galaxy have a subsolar metallicity of 0.6$\pm$0.2Z$\solar$.
Spectroscopy of HII regions in the tail suggest gas metallicities of 0.36-0.53Z$\solar$, similar to the stars in the main body.

9. We propose that IC3418 is a dI which was very recently almost completely stripped of its gas by
ram pressure. 
Neither H$\alpha$ nor HI emission are detected in the main body of the galaxy,
despite the presence of spiral arms and features resembling star forming regions. As its last generation of stars fades and reddens, and substructure
within the stellar disk disperses, the galaxy will presumably become a dE.
Since IC3418 has a color gradient, it could evolve to become one of blue center dEs discussed by \citet{lis06b},
and since it has spiral structure and a disk, it could become one of the disk substructure dEs of \citet{lis06a}.

10. An unresolved source near the galaxy center with M$_{\rm V}$=-9.4 and B-V=0.08 has a velocity which matches the galaxy, and may be a nuclear star cluster. Photometry of other unresolved bright sources in the direction of IC3418 reveal that several of them are too bright to be supergiants in IC3418, so are likely foreground Milky Way stars. Others could be either foreground stars, or supergiants or globular clusters in IC3418.  

11. We propose that the "fireball phenomenon" seen in IC3418 is not the default mode of ram pressure stripping
in the Virgo cluster, but probably arises from very strong ram pressure
which IC3418 is likely experiencing since it is probably close to the cluster core.
Many spiral galaxies in Virgo are known to be experiencing ram pressure
but don't have UV-bright tails or fireballs.
The other galaxies with known UV-bright tails and fireballs are in more massive clusters with stronger 
ram (and thermal) pressure. 

12. We point out that both ram pressure stripping and tidal interactions are required to explain the properties of most dEs.
Ram pressure stripping is decisive for forming a dE since ram pressure stripping without tidal interactions will make a dE,
but a tidal interaction without ram pressure stripping will not make a dE. Some dEs have properties consistent with only ram pressure stripping but no significant tidal interaction. Dwarf galaxies in clusters are expected to experience both ram pressure stripping and tidal interactions, but the timing of the interactions will generally be different.  Ram pressure stripping occurs once upon first cluster infall, and tidal interactions are expected to occur multiple times, both before and after first infall, with tidal damage increasing over time.

\acknowledgements

We are grateful to the staffs of the WIYN and Keck observatories for
their assistance obtaining the data. This research has made use of the GOLD Mine Database \citep{gav03}, and of the NASA/IPAC Extragalactic Database (NED) which is operated by the Jet Propulsion Laboratory, California Institute of Technology, under contract with the National Aeronautics and Space Administration.  PJ acknowledges support from Project M100031203 of the Academy of Sciences of the Czech Republic. We thank the anonymous referee for comments which helped improve the manuscript.

\facility
{\it Facilities: } \rm \facility{Keck }, \facility{WIYN }, \facility{GALEX }

\begin{table}[htpb]
\begin{tabular}{lllccccccc}
\multicolumn{10}{c}{Table 1. Stellar Objects in or toward IC3418} \\
	\hline
Name &  RA (J2000) & DEC (J2000) & R$_{\rm gal}$ & V & B-V & V-R & M$_{\rm V}$  & V$_{\rm helio}$ & Notes \\
 &    &  &   &  (mag) & (mag) & (mag) & (mag) & (km/s) & \\
	\hline
S1  & 12 29 45.32  & 11 24 20.4 & 0.53' & 19.10 & 1.03  & 0.14 & (-12.01) & &  FS,GC?\\
S2  & 12 29 44.55  & 11 24 24.5 & 0.40' & 20.31 & 1.43  & 0.60 & (-10.8) & & FS,GC?\\
S3  & 12 29 44.61  & 11 24 13.7 & 0.33' & 21.99 & 0.38  & 0.17 & (-9.12) & &  FS,SG,GC?\\
S4  & 12 29 43.98  & 11 24 05.9 & 0.17' & 22.29 & 0.81  & 1.13 & (-8.82) & &  FS,SG,GC?\\
S5  & 12 29 43.31  & 11 24 08.9 & 0     & 21.68 & 0.08  & 0.05 & -9.43  & 176$\pm$5 & nucleus? \\
S6  & 12 29 42.58  & 11 24 03.5 & 0.20' & 16.51 & 0.90  & 0.09 & (-14.6) & & FS\\
S7  & 12 29 42.02  & 11 24 02.5 & 0.33' & 24.81 & ---   & 3.09 & (-6.3) & & FS,SG,GC?\\
	\hline
\end{tabular}
\caption{Notes to Table 1:
Offsets measured with respect to S5, at or near the center of IC3418.
Absolute magnitudes if source is in IC3418 at distance of 16.7 Mpc. Given in parentheses for foreground stars or if distance is unknown.
S5 has measured velocity that puts it in IC3418.
FS = foreground star in Milky Way.
SG = supergiant star in IC3418.
GC = globular star cluster in IC3418.
}
\end{table}

\begin{table}[htpb]
\begin{tabular}{lllcccc}
\multicolumn{7}{c}{Table 2. Tail HII Regions in IC3418} \\
	\hline
Name &  RA (J2000) & DEC (J2000) & R$_{\rm gal}$ & H$\alpha$+[NII] Flux  &  V$_{\rm helio}$ & Notes \\
 &    &  &   &    (10$^{-17}$ erg s$^{-1}$ cm$^{-2}$) & (km/s) & \\
	\hline
UV3-H$\alpha$1  & 12 29 46.70 & 11 23 08.3 &  1.33$'$ & $\sim$3 &   236 & a \\
UV3-H$\alpha$2  & 12 29 46.98 & 11 23 04.7 &  1.42$'$ & $\sim$1 &   253 & a  \\
UV5-H$\alpha$1  & 12 29 50.60 & 11 22 50.8 &  2.17$'$ &  16 &   218  &  F3 \\
UV6-H$\alpha$1  & 12 29 50.82 & 11 22 35.0 &  2.40$'$ &  123 &   220  &  K6 \\
UV7-H$\alpha$1  & 12 29 52.77 & 11 22 48.9 &  2.62$'$ &  20 &   293  &  K5 \\
UV9-H$\alpha$1  & 12 29 52.11 & 11 22 04.6 &  2.98$'$ &  70 &   225  &  K4 \\
UV9-H$\alpha$2  & 12 29 51.98 & 11 22 06.1 &  2.93$'$ &  $\sim$2 &   226  & a \\
UV10-H$\alpha$1 & 12 29 52.76 & 11 21 54.2 &  3.22$'$ &   30 &   221  &  K2 \\
UV10-H$\alpha$2 & 12 29 54.19 & 11 22 05.4 &  3.33$'$ & 24 &   --  & b,K3-NE \\
UV10-H$\alpha$3 & 12 29 54.00 & 11 22 02.6 &  3.33$'$ & 10 &  254  &   K3-SW \\
UV10-H$\alpha$4 & 12 29 54.55 & 11 21 44.8 &  3.63$'$ & 10 &  273  &  K1 \\
	\hline
\end{tabular}
\caption{Notes to Table 2:
Names of sources in \citet{fuma11}, for those sources detected by them.
a. For HII regions detected only by spectroscopy, an approximate H$\alpha$ flux has been estimated by scaling the peak line intensity to those of sources detected in both imaging and spectroscopy. Accuracy is a factor of $\sim$2. These faint emission line sources are likely powered by single B stars, or are planetary nebulae.
b. UV10-H$\alpha$2 has no velocity since it was not covered by the spectroscopy.
}
\end{table}

\begin{table}[htpb]
\begin{tabular}{lllcccc}
\multicolumn{7}{c}{Table 3. Background Galaxies in Direction of IC3418's Tail} \\
	\hline
Name &  RA (J2000) & DEC (J2000) & R$_{\rm gal}$ & R  &  z & Notes \\
 &    &  &   &    (mag) &  & \\
	\hline
G1  & 12 29 47.04  & 11 23 31.8 & 1.10' & 20.53 & 0.24 & SDSS J122947.04+112332.3 \\
G2  & 12 29 46.41  & 11 23 11.2 & 1.22' & 21.59 & 0.64 &  SDSS J122946.45+112311.8 \\
G3  & 12 29 50.27  & 11 22 56.8 & 2.08' & 22.75 & 0.68 & \\
G4  & 12 29 50.97  & 11 22 48.3 & 2.31' & 21.58 & 0.37 & \\
G5  & 12 29 52.22  & 11 22 03.3 & 3.02' & 22.33 & 0.23 &  \\
	\hline
\end{tabular}
\caption{Notes to Table 3:
Offsets (R$_{\rm gal}$) measured with respect to S5, at or near the center of IC3418.
}
\end{table}

\end {document}